\documentclass[12pt]{oxarticle}
\pdfoutput=1

\usepackage{graphicx, float, array, xspace, amscd, amsmath, amsthm, amssymb, latexsym, relsize, bbm}
\usepackage[letterpaper, left=1.7in, right=0.3in, top=0.8in, bottom=1.2in]{geometry}
\usepackage[bbgreekl]{mathbbol}
\usepackage{amsfonts, tikz-cd}
\usepackage[T1]{fontenc}

\DeclareSymbolFontAlphabet{\mathbb}{AMSb}
\DeclareSymbolFontAlphabet{\mathbbl}{bbold}

\usepackage[sort&compress, comma, square, numbers]{natbib}
\usepackage[all,cmtip]{xy}
\usepackage{color}
\definecolor{MyDarkBlue}{rgb}{0.15,0.25,0.45}
\usepackage[linktocpage=true]{hyperref}
\hypersetup{colorlinks=true, citecolor=blue, linkcolor=blue, urlcolor=blue, pdfauthor={}, pdftitle={},breaklinks=true}

\let\SS=\S 

\renewcommand{\#}{^{\sharp}}

\newcommand{\Het}{\hbox{\sf Het}}

\newcommand{\Ieta}{\mathbbl{\bbeta}}
\newcommand{\LC}{\text{\tiny LC}}
\newcommand{\Hu}{{\text{\tiny H}}}
\newcommand{\Bi}{{\text{\tiny B}}}

\newcommand{\Jsh}{{J^\sharp{}}}
\newcommand{\Ch}{{\rm Ch}}

\renewcommand{\sb}{{\overline{\sigma}}}
\newcommand{\chib}{\overline{\chi}}

\newcommand{\kb}{{\overline{ \kappa}}}
\newcommand{\rb}{{\overline{ r}}}

\newcommand{\Ob}{{\overline{ \Omega}}}

\newcommand{\bfB}{{\bf B}}

\newcommand{\bfE}{{\bf E}}
\newcommand{\SO}{{\rm SO}}

\newcommand{\w}{{\,\wedge\,}}
\newcommand{\wt}{\widetilde}

\newcommand{\so}{{\mathfrak{so}}}
\newcommand{\fD}{{\mathfrak{D}}}
\newcommand{\fDb}{{\overline\fD}}

\newcommand{\delslash}{\ensuremath \raisebox{0.025cm}{\slash}\hspace{-0.23cm} \del}
\newcommand{\Hslash}{\hspace{0.1cm}\ensuremath \raisebox{0.03cm}{\slash}\hspace{-0.30cm} H}

\newcommand{\Fslash}{\hspace{0.1cm}\ensuremath \raisebox{0.025cm}{\slash}\hspace{-0.28cm} F}

\newcommand{\half}{\frac{1}{2}}

\def\CS{{\text{CS}}}

\newcommand{\ab}{{\overline\alpha}}
\newcommand{\bb}{{\overline\beta}}

\newcommand{\db}{{\overline\delta}}

\renewcommand{\a}{\alpha}
\renewcommand{\b}{\beta}
\newcommand{\g}{\gamma}\newcommand{\G}{\Gamma}
\renewcommand{\d}{\delta}\newcommand{\D}{\Delta}
\newcommand{\e}{\epsilon}\newcommand{\ve}{\varepsilon}
\newcommand{\z}{\zeta}

\renewcommand{\th}{\theta}\newcommand{\Th}{\Theta}\newcommand{\vth}{\vartheta}

\renewcommand{\k}{\kappa}
\renewcommand{\l}{\lambda}\renewcommand{\L}{\Lambda}
\newcommand{\m}{\mu}
\newcommand{\n}{\nu}
\newcommand{\x}{\xi}

\newcommand{\p}{\pi}
\renewcommand{\r}{\rho}
\newcommand{\s}{\sigma}\renewcommand{\S}{\Sigma}
\renewcommand{\t}{\tau}

\newcommand{\Ph}{\Phi}\newcommand{\vph}{\varphi}
\newcommand{\ch}{\chi}
\newcommand{\Ps}{\Psi}
\renewcommand{\o}{\omega}\renewcommand{\O}{\Omega}

\DeclareFontFamily{OT1}{pzc}{}
\DeclareFontShape{OT1}{pzc}{m}{it}{<-> s * [1.200] pzcmi7t}{}
\DeclareMathAlphabet{\mathpzc}{OT1}{pzc}{m}{it}

\newcommand{\cA}{\mathcal{A}}
\newcommand{\ccB}{\mathpzc B}

\newcommand{\ccD}{\mathpzc D}
\newcommand{\ccE}{\mathpzc E}

\newcommand{\ccK}{{\mathpzc K}\,}
\newcommand{\cL}{\mathcal{L}}\newcommand{\ccL}{\mathpzc L}
\newcommand{\cM}{\mathcal{M}}

\newcommand{\cO}{\mathcal{O}}

\newcommand{\cR}{\mathcal{R}}

\newcommand{\cT}{\mathcal{T}}\newcommand{\ccT}{\mathpzc T}
\newcommand{\ccU}{\mathpzc U}

\newcommand{\cZ}{\mathcal{Z}}\newcommand{\ccZ}{\mathpzc Z}

\newcommand{\ccZb}{{\overline \ccZ}}
\newcommand{\cZb}{{\overline \cZ}}

\DeclareFontFamily{U}{bbold}{}
\DeclareFontShape{U}{bbold}{m}{n}
 {  <-5.5> s*[1.05] bbold5
    <5.5-6.5> s*[1.05] bbold6
    <6.5-7.5> s*[1.05] bbold7
    <7.5-8.5> s*[1.05] bbold8
    <8.5-9.5> s*[1.05] bbold9
    <9.5-11.5> s*[1.05] bbold10
    <11.5-16> s*[1.05] bbold12
    <16-> s*[1.05] bbold17
 }{}

\newcommand{\IA}{\mathbbl{A}}
\newcommand{\IB}{\mathbbl{B}}
\newcommand{\IC}{\mathbbl{C}}
\newcommand{\Id}{\mathbbl{d}}

\newcommand{\IF}{\mathbbl{F}}
\newcommand{\Ig}{\mathbbl{g}}
\newcommand{\IH}{\mathbbl{H}}

\renewcommand{\IJ}{\mathbbl{J}}
\newcommand{\IK}{\mathbbl{K}}
\newcommand{\IL}{\mathbbl{L}}

\newcommand{\IR}{\mathbbl{R}}
\newcommand{\Is}{\mathbbl{s}}

\newcommand{\IU}{\mathbbl{U}}
\newcommand{\Iv}{\mathbbl{v}}
\newcommand{\IW}{\mathbbl{W}}\newcommand{\Iw}{\mathbbl{w}}
\newcommand{\IX}{\mathbbl{X}}
\newcommand{\IY}{\mathbbl{Y}}
\newcommand{\IZ}{\mathbbl{Z}}

\newcommand{\Ibeta}{\mathbbl{\bbbeta}}
\newcommand{\IGamma}{\mathbbl{\Gamma}}

\newcommand{\ITheta}{\mathbbl{\Theta}}

\newcommand{\Ichi}{\mathbbl{\bbchi}}

\newcommand{\Iomega}{\mathbbl{\hskip1pt\bbomega}}

\newcommand{\inbar}{\vrule height 0.43em width 0.048em}
\newcommand{\Idel}{\hbox{$\partial\hspace{-0.38em\inbar\hspace{0.3em}}$}}
\newcommand{\Idelb}{\hbox{$\bar\partial\hspace{-0.38em\inbar\hspace{0.3em}}$}}

\newcommand{\IcA}{\hbox{$\cA$\hspace{-0.67em}\raisebox{-0.05em}{\vrule width 0.12em height 0.06em} \hspace{-0.4em}\raisebox{0.35pt}{$\scriptstyle\cA$}\hspace{-0.6em}\raisebox{0.1pt}{$\scriptstyle\cA$}}}

\newcommand{\Inabla}{\nabla\raisebox{0.44ex}{\hspace{-1.45ex}\text{\smaller[2]$\nabla$}}}

\newcommand{\one}{\mathbbm{1}}

\newcommand{\Deth}{\text{\DH}\hskip1pt}
\newcommand{\Dethsharp}{\Deth^\sharp}
\newcommand{\Dethsharpb}{\overline{\Deth^\sharp}}
\newcommand{\deth}{\eth} 
\newcommand{\dethb}{\overline{\eth}}
\newcommand{\dethsharp}{\eth^\sharp}
\newcommand{\dethsharpb}{\overline{\eth^\sharp}}

\font\eightrm=cmr8 at 8pt
\font\csc=cmcsc10

\newcommand{\beq}{\begin{equation}}
\newcommand{\eeq}{\end{equation}}
\newcommand{\beqnn}{\begin{equation*}}
\newcommand{\eeqnn}{\end{equation*}}
\newcommand{\bea}{\begin{eqnarray}}
\newcommand{\eea}{\end{eqnarray}}
\newcommand{\bean}{\begin{eqnarray*}}
\newcommand{\eean}{\end{eqnarray*}}

\newcommand{\fref}[1]{Figure~\ref{#1}}
\newcommand{\tref}[1]{Table~\ref{#1}}
\newcommand{\sref}[1]{\SS\ref{#1}}

\newcommand{\pd}[2]{\frac{\partial #1}{\partial #2}}

\newcommand{\norm}[1]{\left\| #1\right\|}

\newcommand{\ii}{\text{i}}

\newcommand{\place}[3]{\vbox to0pt{\kern-\parskip\kern-7pt
                             \kern-#2truein\hbox{\kern#1truein #3}
                             \vss}\nointerlineskip}
\newcommand{\smallfrac}[2]{\frac{\scriptstyle #1}{\scriptstyle #2}}

\DeclareFontFamily{U}{wncy}{}
\DeclareFontShape{U}{wncy}{m}{n}{<->wncyr10}{}
\DeclareSymbolFont{mcy}{U}{wncy}{m}{n}
\DeclareMathSymbol{\sha}{\mathord}{mcy}{"58}

\newcommand{\eu}[1]{{\mathfrak #1}}

\newcommand{\capt}[3]{\parbox{#1}{\renewcommand{\baselinestretch}{1.0}
                                                           \caption{\label{#2}\small\it #3}}}

\newcommand{\del}{\partial}
\newcommand{\delb}{\overline{\partial}}

\newcommand{\lb}{{\overline\lambda}}
\newcommand{\nb}{{\overline\n}}
\newcommand{\mb}{{\overline\m}}

\newcommand{\A}{\cA}

\newcommand{\dd}{\text{d}}

\newcommand{\cym}{Calabi-Yau manifold\xspace}

\newcommand{\K}{K\"ahler\xspace}

\def\im{{\rm im ~}}
\def\ker{{\rm ker ~}}
\newcommand{\+}{\phantom{-}}

\newcommand{\tr}{\text{Tr}\hskip2pt}

\newcommand{\tb}{{\overline{\tau}}}

\newcommand{\ap}{{\a^{\backprime}\,}}
\renewcommand{\sb}{{\overline{\sigma}}}
\renewcommand{\rb}{{\overline{\rho}}}
\renewcommand{\=}{\;=\;}

\makeatletter
\g@addto@macro\bfseries{\boldmath}
\makeatother

\newcommand{\citeI}{\cite{Candelas:2016usb}\xspace}
\renewcommand{\baselinestretch}{1.1}
\numberwithin{equation}{section}
\proofmodefalse
\begin{document}
\pagestyle{empty}      
\ifproofmode\underline{\underline{\Large Working notes. Not for circulation.}}\else{}\fi

\begin{center}
\null\vskip0in
{\Huge The Universal Geometry of Heterotic Vacua \\[0.2in]}
{\csc Philip Candelas$^{*\,1}$, Xenia de la Ossa$^{*\,2}$, Jock McOrist$^{\dagger \,3}$\\[-1pt]
and\\[-1pt]
Roberto Sisca$^{\dagger \,4}$\\[0.3in]}

{\it $^*$Mathematical Institute\hphantom{$^1$}\\
University of Oxford\\
Andrew Wiles Building\\
Woodstock Road, Radcliffe Observatory Quarter\\
Oxford, OX2 6GG, UK\\[2ex]

 $^\dagger$Department of Mathematics\hphantom{$^2$}\\
University of Surrey\\
Guildford, GU2 7XH, UK}

\footnotetext[1]{{\tt candelas@maths.ox.ac.uk} \hfil
$^2\,${\tt delaossa@maths.ox.ac.uk} \hfil
$^3\,${\tt j.mcorist@surrey.ac.uk}\\[2pt]
\hspace*{6.21cm}$^4\,${\tt r.sisca@surrey.ac.uk}}
\vskip30pt
{\bf Abstract\\}
\end{center}
\vskip-12pt
\renewcommand{\baselinestretch}{0.9}
We consider a family of perturbative heterotic string backgrounds. These are complex threefolds $X$ with $c_1{\=}0$, each with a gauge field solving the Hermitian Yang-Mill's equations and compatible $B$ and $H$ fields that satisfy the anomaly cancellation conditions. Our perspective is to consider a geometry in which these backgrounds are fibred over a parameter space. If the manifold $X$ has coordinates $x$, and parameters are denoted by $y$, then it is natural to consider coordinate transformations $x\to \tilde{x}(x,y)$ and $y\to \tilde{y}(y)$. Similarly,  gauge transformations of the gauge field and $B$ field also  depend on both $x$ and $y$. In the process of defining deformations of the background fields that are suitably covariant under these transformations, it turns out to be natural to extend the gauge field $A$ to a gauge field $\IA$ on the extended $(x,y)$-space. Similarly, the $B$,~$H$, and other fields are also extended. The total space of the fibration of the heterotic structures is the Universal Geometry of the title. The extension of gauge fields has been studied  in relation to Donaldson theory and monopole moduli spaces. String vacua furnish a richer application of these ideas. One advantage of this point of view is that previously disparate results are unified into a simple tensor formulation. In a previous paper, by three of the present authors, the metric on the moduli space of heterotic theories was derived, correct through $\cO(\ap)$, and it was shown how this was related to a simple \K potential. With the present formalism, we are able to rederive the results of this previously long and involved calculation, in less than~a~page.
\newgeometry{left=1.5in, right=0.5in, top=0.75in, bottom=0.8in}
%
\newpage
{\baselineskip=10pt\tableofcontents}
\restoregeometry
%
%
\newpage
\setcounter{page}{1}
\pagestyle{plain}
\renewcommand{\baselinestretch}{1.3}
\null\vskip-10pt
\rightline{Nullum autem est, nisi quod animus ex se sibi invenit.}
\rightline{\it There can be no good except for what the soul discovers for itself within itself.}
\rightline{Seneca the Younger}
\vskip25pt
\section{Introduction}
\vskip-10pt
\subsection{Preamble}
\vskip-10pt
Heterotic geometry is the geometry associated with the moduli space of a heterotic vacuum of string theory. The geometrical background, associated with a vacuum, is understood, at large volume, as $\IR^{1,3}{\times} X$, where $X$ is a complex 3-dimensional manifold with vanishing first Chern class. This geometry is endowed with a holomorphic vector bundle $\ccE\to X$, admitting a connection $A$ that satisfies the Hermitian Yang-Mills equations 
\beq
g^{\m\nb} F_{\m\nb} \= 0~.
\label{eq:HYM}\eeq
Also important are  background values for the Kalb--Ramond field $B$ and its field strength~$H$, which satisfy appropriate field equations and anomaly cancellation conditions. Heterotic geometry is the analogue of the special geometry of Type II vacua. 

We term the tuple $([X,\o,\O], [\ccE,A], [\ccT_X,\Th], H)$ a heterotic structure and label it by $\Het$. The data of the heterotic structure includes the connections $A$ and $\Th$ on the bundles $\ccE$ and $\ccT_X$ respectively as well as the hermitian form $\o$ and the complex structure $J$ of $X$, or equivalently the holomorphic 
$(3,0)$-form $\O$. 

The metric on the moduli space of heterotic supergravity metric was computed, correct to~$\cO(\ap)$ (i.e. the error is $\cO(\ap^2)$), in \cite{Candelas:2016usb} by a dimensional reduction of heterotic supergravity. This metric has to be \K as a consequence of supersymmetry. It should not be surprising, therefore, that verifying that the moduli space is in fact \K requires taking into account the relations between $H$, the connection on the bundle $\ccE$, and the hermitian form $\o$ on $X$, since these relations follow from both the anomaly cancellations condition and the requirement of supersymmetry.

Let us recall these essential conditions. The anomaly relation yields a modified Bianchi identity for $H$. 
\beq
\dd H ~=- \frac{\ap}{4} \Big( \tr (F^2) - \tr (R^2)\, \Big)~,
\label{eq:Anomaly0}\eeq
while the supersymmetry relation takes the form\footnote{The right hand side of this relation is often written in the form $\ii (\del -\delb)\o$. However we prefer to write this in ``real form'' since this makes the calculation of the derivative of the relation with respect to the complex structure parameters more transparent.}
\beq
 H \= \dd^c \o~,\quad \dd^c \o \= \frac{1}{3!}J^mJ^n J^p (\dd \o)_{mnp}~.
\label{eq:SusyRelation0}\eeq
In the above equations $R$ is the curvature two-form. We denote by $x^m$ the real coordinates of $X$ and the holomorphic coordinates by $(x^\m, x^\nb)$. The vector-valued form 
$J^m {\=} J_n{}^m \dd x^n$ is a 1-form constructed from the complex structure. In the following we will generally omit the wedge product symbol `$\wedge$' between forms, unless doing so would lead to ambiguity.

These equations already imply that the moduli space has a recondite character, since the deformations of $F$, $\o$ and $H$ are intricately related. By contrast to the case of type II vacua, where the roles of the complex structure parameters and the \K class parameters are strictly separated, there seems to be no useful distinction, in the heterotic context, between what are conventionally labelled the complex structure moduli, hermitian moduli and bundle moduli. 

The deformations of a heterotic structure, within a given topological class, correspond to the points of the moduli space $M$, which is itself a complex manifold. This has real coordinates~$y^a$ and complex coordinates  
$(y^\a,y^\bb)$. 

We think of the heterotic structures as fibred over $M$ and denote the total space by $\ccU$. This is the universal bundle of the title. The fibration $\ccU$ is of course distinct from the holomorphic vector bundle $\ccE$, which is part of the heterotic data of each fiber of $\ccU$. Thus we have the diagram
\beq\label{eq:HeteroticFamily}
\begin{tikzcd}
\Het \arrow[r]  & \ccU \arrow[d]\\
& M
\end{tikzcd}~.
\eeq

The purpose of this article is to show that considering this structure is worthwhile. Before entering into technical matters it may be helpful to indicate why this might be expected to be the case. To start, consider for example the deformation of a manifold, which is part of our data. In general relativity one often thinks of a 
three--geometry that evolves in time. We think of time as a parameter which governs the evolution. More generally, a manifold may depend on a number of parameters so 
there are a number of `times'. A minor, but still counter-intuitive matter to a physicist, is that whereas the time axis almost always proceeds upwards in diagrams, in the fibre bundle language the parameter space is the base
of the fibration and is invariably drawn horizontally while the different manifolds are the fibers and are drawn vertically. Thus time evolves sideways.

Let us take another analogy with non-relativistic electrodynamics. The fields $\bfE$ and $\bfB$ reside in 
$\IR^3$ and evolve in time. Of course, if we pass to the four-dimensional description of relativistic dynamics, time is included as one of the coordinates.  The vectors of $\IR^3$, are now no longer covariant under the enlarged symmetry group of $\SO(1,3)$. Instead 
$\bfE$ and $\bfB$ come together into a two-form $F_{mn}$ that transforms covariantly under the larger group. 

To see how a `larger group' arises in the present context, consider the Yang-Mills gauge field~$A$, that is present in a heterotic vacuum. It is subject to gauge transformations
\beq
A\to\; ^\Ph\!A \= \Ph A \Ph^{-1} - \dd\Ph\,\Ph^{-1}~.
\label{eq:Atransf}\eeq
Since we wish to consider deformations of the gauge potential, the gauge function $\Ph$ is naturally a function of both the coordinates of $x$ of $X$ and the parameters $y$ of $M$. It is convenient to define a covariant derivative $D_a A$, which transforms homogeneously under gauge transformation. In order to do this in a way which takes into account the parameter dependence of the gauge  transformations, one introduces also a connection 
$\L {\=} {\L}_a\, \dd y^a$ on the moduli space that transforms in a manner parallel to $A$ 
\beq
\L\to\; ^\Ph\L_a\= \Ph \L_a \Ph^{-1} - \del_a\Ph\,\Ph^{-1}~.
\label{eq:Ltransf}\eeq
The covariant derivative contains the two connections $A$ and $\L$ 
\beq\label{eq:covariantderivative1}
D_a A \= \del_a A - \dd_A \L_a~,
\eeq
where $\dd_A$ denotes a covariant outer derivative operator
\beq
\dd_A \L_a \= \dd\L_a + [A,\,\L_a]~.
\notag\eeq
Under a gauge transformation we see that the covariant derivative $D_a A$ is indeed covariant
\beq
D_a A \to {}^\Ph D_a A \= \Ph\, D_a A\, \Ph^{-1}~.
\notag\eeq

There are certain useful identities, that can be derived in a straightforward manner that relate to this covariant derivative. Consider, for example, these two
\beq
 \dd_A(D_a A)\= D_a F~, ~~~\text{and}~~~\quad [D_a, D_b] A ~=-\dd_A \IF_{ab}~,
\label{eq:CovDiffIds}\eeq
where $\IF_{ab}{\=}\del_a\L_b - \del_b\L_a + [\L_a, \L_b]$. 

Let us combine the connections $A$ and $\L_a$ into the single connection
\beq
\IA\= A + \L_a \dd y^a~.
\notag\eeq
and form from this the associated field strength
\beq
\IF \= \Id \IA + \IA^2~,~~~\text{with}~~~\Id\= \dd + \dd y^a \del_a~.
\notag\eeq
With this definition $\IF$ is covariant. Moreover: 
\begin{itemize}
\item $\IF_{ab}$ is as defined above\bigskip
\item $D_a A{\=}\IF_{\!\!am}\dd x^m$, so the covariant derivative is none other than the mixed component of the covariant field strength $\IF$. 
\item Furthermore, it is easy to check that the somewhat subtle, but useful, identities \eqref{eq:CovDiffIds} are the $(a,m,n)$ and $(a,b,m)$ components of the Bianchi identity
\beq
\Id_{\IA}\IF \= 0~.
\notag\eeq
\end{itemize}

The spin connection $\Th$ plays a role with respect to the tangent bundle $\ccT_X$ that is closely parallel to the role played by $A$ with respect to the vector bundle $\ccE$. Under a Lorentz transformation $\Psi$  the transformation law for $\Th$ is 
\beq
\Th \to \Psi \Th\Psi^{-1} - \dd\Psi\Psi^{-1} ~.
\label{eq:ThGaugeTransf}\eeq
This has field strength $R {\=} \dd \Th + \Th^2$, which is holomorphic and obeys the Hermitian--Yang--Mills equation \eqref{eq:HYM}. So, for this connection, it is natural to define an extended connection and field strength analogous to those above
\beq\label{eq:IR}
\ITheta\= \Th +\Xi_a\dd y^a~,~~~\IR \= \Id\ITheta + \ITheta^2~,
\eeq
where $\Xi_a$ is the analogue of the connection $\L_a$. 

The point that is being made is that there is no clean separation between the coordinates $x$ of the manifold $X$ and the parameters $y$. We have seen that the gauge functions $\Ph$ and $\Ps$ are naturally functions of both $x$ and $y$. Furthermore, it is natural that the diffeomorphisms of $X$ should involve the parameters $y$. That is, we are allowed diffeomorphisms of the form
\beq 
x \;\to\; \wt x(x,y)~;~~~y \;\to\; \wt y(y)~. 
\label{eq:diffeomorphisms}\eeq
This, taken together with the dependence of the gauge functions on both $x$ and $y$ constitutes the `larger group'.

We are not the first to consider a universal bundle, in this sense. What has been said so far in this context applies to the deformation theory of gauge fields. It was originally considered by Atiyah--Singer \cite{AtiyahSinger:1984}, and further exploited in connection with Donaldson Theory \cite{Donaldson1,Donaldson2,Witten:1990} and its relation with five-branes \cite{Harvey:1991hq}. Gauntlett adapted the approach of \cite{Harvey:1991hq}  to the moduli space of BPS monopoles \cite{Gauntlett:1993sh}. The covariant derivatives, analogous to those constructed above, were observed to have an interpretation as the extended field strength of a universal bundle. Heterotic string theory, however, furnishes a richer, or depending on one's point of view, more complicated structure. In particular the Kalb-Ramond field $B$ is a potential of an unconventional type for the $H$ field since, technically, it is a 2-gerbe (see \cite{Hitchin:Gerbe} for a broad introduction to gerbes). The field  $B$ also transforms under gauge transformations and it is convenient also to try to define a covariant derivative for this field. It is not possible to define one that transforms homogeneously, so we settle for a derivative whose transformation law is as closely parallel as possible to the transformation law for $B$ itself. 

As we will see later, a quantity $\ccB_a$ arises that is closely related to the covariant derivative, and does have the property that it transforms homogeneously. The quantity $\ccB_a$ is of additional interest since the combination $\ccB_\a + \ii \fD_\a\o$, where $\a$ is a holomorphic index and
$\fD_\a\o$ is a certain covariant derivative of the hermitian form $\o$, take over the role that the derivatives of the complexified \K form, $\del_\a (B+\ii\o)$, play in special geometry. We will review this now, but the upshot is that extended quantities that include $\IB$ and $\IH$ play an important role~also.

The field strength $H$ is related to $B$ by the relation
\beq
H \= \dd B - \frac{\ap}{4}\Big(\CS[A] - \CS[\Th]\Big)~,
\label{Hdef}\eeq
where $\CS$ denotes the Chern--Simons three-form
\beq
\CS[A] \= \tr\!\left(A\dd A +\frac{2}{3}\, A^3\right) ~.
\notag\eeq
The Chern--Simons forms transform under gauge transformation and so does $B$, with the transformation law for $B$ chosen to ensure that $H$ is gauge invariant. This transformation law is 
\beq
{}^{\Phi,\Psi} B \= B + \frac{\ap}{4}\Big( \tr \big(Y\! A - Z \Th\big) +  U\! - W \Big)~,
\label{eq:BTransf2} \eeq
with 
\beq
Y\=\dd\Phi \Phi^{-1}~,\qquad Z\=\dd\Psi \Psi^{-1}~,
\notag\eeq
and $U$ and $W$ are such that $\dd U {\=} \frac{1}{3} \tr (Y^3)$ and $\dd W {\=} \frac{1}{3} \tr (Z^3)$. 

As $H$ is gauge invariant its variation with respect to the parameters can simply be given as a partial derivative. In this way we arrive at a relation of the form
\beq
\del_a H \= \dd \ccB_a - \frac{\ap}{2} \tr \big(D_a A\, F\big) + \frac{\ap}{2} \tr \big( D_a \Th R \big)~.
\label{eq:Hderiv}\eeq
This relation identifies a gauge invariant quantity $\ccB_a$ that is defined up to the addition of a $\dd$-closed form.

Let us define extended forms of $B$ and $H$ that are related by
\beq
\IH \= \Id \IB - \frac{\ap}{4} \Big(\CS[\IA] - \CS[\ITheta]\Big)~,~~~\text{where}~~~
\CS[\IA] \= \tr\!\left(\IA\,\Id \IA +\frac{2}{3}  \IA^3\right)~.
\label{eq:IHdef2}\eeq
It is pleasing that the important quantity $\ccB_a$ turns out to be a mixed component of the gauge invariant tensor $\IH$
\beq
\ccB_a \= \frac12 \IH_{amn} \dd x^m \dd x^n~.
\notag\eeq
We take $\IH$ to satisfy an extended supersymmetry relation  and a Bianchi identity
\begin{equation}\notag
 \IH \= \Id^c\Iomega ~,\quad \Id\IH~=-\frac{\ap}{4}\Big(\tr{\IF^2}-\tr{\IR^2}\Big)\ ,
\end{equation}  
whose mixed components give important relations \eqref{susyoncF}, \eqref{susyoncF2} and \eqref{secondorderom} among the heterotic moduli. 

So far, we have discussed the consequence of allowing the gauge functions $\Ph$ and $\Ps$ to be functions of both $x$ and $y$.
In order to discuss the extension of vectors and tensors to the bigger space we have to take into account the freedom expressed by the second of equations \eqref{eq:diffeomorphisms}. We are led to introduce a covariant basis of forms and a corresponding dual basis of vectors
\begin{align*}
e^m &\= \dd x^m + c_a{}^m\dd y^a~, & e^a &\= \dd y^a~, \\[5pt]
e_m &\= \del_m~, & e_a~,&\= \del_a - c_a{}^m \del_m~.
\end{align*}
The quantity $c^m{\=}c_a{}^m \dd y^a$ is a connection which transforms, under $x\to \tilde{x}(x,y)$, in the form
\beq
c^{\widetilde{m}} \= \pd{x^{\widetilde{m}}}{x^n}\, c^n - \pd{x^{\widetilde{m}}}{y^b}\, \dd y^b~,
\label{eq:ctransf}\eeq
and this ensures that the forms $e^m$ and vectors $e_a$ transform as expected
\beq
e^{\widetilde{m}} \= \pd{x^{\widetilde{m}}}{x^n}\, e^n~,~~~
e_{\widetilde{a}}  \= \pd{y^b}{y^{\widetilde{a}}}\, e_b~.
\notag\eeq
With these basis forms, we write the extended vector potential, for example, as
\beq
\IA \= A_m e^m + A^\sharp_a\dd y^a~,~~~\text{with}~~~A^\sharp_a \= \L_a - A_m c_a{}^m~. 
\notag\eeq
We will introduce a pair of covariant derivatives $\Deth$ and $\Dethsharp_a$ for the 
connection~$c_a{}^m$. The operator $\Deth$ covariantises the de Rham operator $\dd$ along the manifold $X$, and so within $\IX$ defines a fibre-wise cohomology;  while the operator $\Dethsharp$ describes how tensors on $X$ change under a change in parameters. In order to describe the variation of gauge dependent quantities we will also introduce a covariant derivative $\fD_a$, which is covariant with respect to both diffeomorphisms and gauge symmetries. 

Inevitably, we are led to define covariant derivatives that decompose with respect to the fibre $X$ and the base $M$, and again, when we come to discuss the complex structure, with respect to the complex structures of $X$ and $M$. Thus we will have a fourfold decomposition of the derivatives, which we require to be covariant. By covariant we mean covariance with respect to the diffeomorphisms \eqref{eq:diffeomorphisms}, which is more general than the situation considered in \citeI. This allowed the gauge functions to be functions of both $x$ and $y$, but which only envisioned diffeomorphisms
$(x,\,y)\to (\tilde{x}(x),\,\tilde{y}(y))$. The freedom to make transformations $x\to\tilde{x}(x,y)$ requires the introduction of the connection $c_a{}^m$ and forces a redefinition of the derivatives. In an attempt to make this transition as easy as possible we have summarised these in \sref{sec:derivativetab}.
 
The connection $c^m$ can be identified with the cross term in the `minimal' extended metric. We have two metrics that arise naturally:
the metric $g_{mn}$ on the manifold $X$ and the metric $g^\sharp_{ab}$ on the moduli space $\cM$. If we combine these into a `minimal' extended metric
\beq
\Id\Is^2 \= g_{mn} e^m e^n + g^\sharp_{ab} \dd y^a \dd y^b~,
\notag\eeq
and write this out in terms of the basis forms $\dd x^m$ and $\dd y^a$ we find the cross term
\beq
\Ig_{ma} \= g_{mn} c_a{}^n~.
\notag\eeq

The connection $c^m$ appears also in an interesting way in relation to the variation of the complex structure of $X$. The first order variation of the complex structure is recorded in a form $\D_\a{}^\m{\=}\D_{\a\nb}{}^\m \dd x^\nb$ which is defined by
\beq
\D_\a{}^\m \= \del_\a \dd x^\m \Big|^{(0,1)}
\notag\eeq
It will also be shown later that $\D_\a{}^\m$ is also related to $c^\m$ by
\beq
\D_{\a\,\nb}{}^\m \= -\del_\nb c_\a{}^\m~.
\label{eq:extrinsic1}\eeq

The connection $c_a{}^m$ also plays an important geometrical role in relation to the fibration $\IX$, in that it determines an almost product structure $\IL$ that provides a splitting of the tangent space of the fibration $\cT_\IX$ into vertical and horizontal subspaces. Being a fibration, $\IX$ naturally encodes a vertical projection $\IX\longrightarrow M$. A horizontal structure, equivalent to a local choice of $c_a{}^m$, is not invariantly defined. The freedom inherent in choice of $c_a{}^m$ corresponds precisely to the freedom to make coordinate transformations as in \eqref{eq:diffeomorphisms}.

\subsection{Outline of the article}
\vskip-10pt
In the body of the article we give a detailed discussion of the points outlined above.

Within the fibration $\ccU$ lies the fibration of the manifold $X$ over M, 
\beq
\begin{tikzcd}
X \arrow[r]  & \IX \arrow[d]\\
& M
\end{tikzcd}~.
\notag\eeq
This is the natural context in which to discuss the Ehresmann connection --- equivalently, the projection~$\p$ --- the metric $\Ig$ and complex structure $\IJ$ for the extended space. This is the subject of 
\sref{sec:fibrationccX}. The connection $c_a{}^m$ allows us to restrict $\Ig$ and $\IJ$ to fibres covariantly and, when this is done, they are identified with the metric $g$ and complex structure $J$ on~$X$. Furthermore, using $\p$ we can also project $\Ig$ and $\IJ$ to the moduli space metric $g^\sharp$ and complex structure $J^\sharp$.   

In section 2, we describe the differential calculus of $\IX$ and its relation to deformations. For example,  we show that the covariant derivatives  such as \eqref{eq:covariantderivative1} are identified as Lie derivatives acting tensors on $\IX$. This leads to an interpretation of deformations with flows on $\IX$.  

In section 3, we start to see the profits of our labour. We introduce on $\IX$ extensions of the connections $A$ and $\Th$, denoted $\IA$ and $\ITheta$ respectively, which allows to discuss the extended symmetry groups mentioned above. The fields $\IA$ and $\ITheta$ are holomorphic connections for the vector bundle $\ccU\to \IX$.  Moreover, we define  the extensions of $\omega$ and $H$, denoted  $\Iomega$ and $\IH$ respectively, and suppose a relation $\IH = \Id^c \Iomega$, as the extension of the supersymmetry 
      relation~\eqref{eq:SusyRelation0}.  

Surprisingly, this relation together with  its Bianchi identity, encapsulate in a simple pair of tensor equations,  a set of long and otherwise complicated  equations relating covariant derivatives which were crucial to the derivation of the \K moduli space metric in \citeI. This is similar to how the laws of electrodynamics when viewed relativistically are unified into a simple tensor equation.

In section 4, we illustrate a utility of $\IX$ by showing  how the curvature $\IR$ in \eqref{eq:SusyRelation0} can be used to compute the covariant derivative $D_\a \Th$  in terms of the complex structure moduli $\D_\a{}^\m$ and hermitian moduli $\fD_\a \o^{1,1}$ to zeroth order in $\ap$. We then use this to compute the last term in  the moduli space metric $g^\sharp$ derived in \citeI to be 
\beq
\newcommand{\negskip}{\hskip-4pt}
\begin{split}\label{eq:modulimetric}
\dd s^{\sharp\,2} &=  2g^\sharp_{\a\bb} \,\dd y^\a  \otimes  \dd y^\bb~; \\[10pt]
g^\sharp_{\a\bb} 
&= 
\frac{1}{V}\hskip-2pt\int_X\hskip-2pt\left\{\D_\a{}^\m \star \D_\bb{}^\nb \, g_{\m\nb} + 
\frac{1}{4}\cZ_\a \star \cZb_\bb + 
\frac{ \ap}{4} \tr \big( \fD_\a A \star \fD_\bb A \big) \right.\\
&\hskip4.3cm\left. +  \frac{\ap}{2} \!\left( \D_{\a\,\mb\nb} \D_{\bb \r\s}+ 
\fD_\a \o_{\r\mb} \,\fD_\bb \o_{\s\nb} \right) R^{\mb\r\nb\s} 
 \right\}~,
\end{split}\raisetag{55pt}
\eeq
which generalises an expression in \cite{Anguelova:2010ed} to include all the moduli. 

In section 5, we put all of this together to show how to derive the moduli space metric $g^\sharp_{\a\bb}$ from its \K potential in a concise way, which simplifies much of the analysis~of~\citeI. 
\subsection{Some notation and terminology}
\label{sref:Notation}
\vskip-10pt
It is useful to summarise some notation and terminology that we will introduce later.
\begin{itemize}
 \item Tangibility $[p,q]$ means the form has $p$ legs along the moduli space $M$ and $q$ legs along the fibre $X$. Our convention is that legs along the moduli space are written first.  
\smallskip  
 \item The corpus of a form is the part with all legs along the fibre $X$. The animus consists of all remaining components of the form, and these are distinguished by a $\sharp$ superscript.
\end{itemize}

\begin{table}[H]
\begin{center}
\setlength{\extrarowheight}{3pt}
\begin{tabular}{|m{6cm}|c|c|}
\hline
\hfil Coordinates &~~Real indices~~ & ~~Complex indices~~\\[3pt]
\hline\hline
~Total space $\IX$\hfill $u^P$~ & $P,Q,R,S$ & \\[3pt]
\hline
~Base manifold $M$\hfill $y^a$~ & $a,b,c,d$ & $\a,\b, \g$ \\[3pt]
\hline
~Fibre manifold $X$\hfill $x^m$~ & $m,n,k,l$ & $\m,\n,\k,\l$ \\[3pt]
\hline
\end{tabular}
\capt{6.0in}{tab:coords}{The coordinates and indices for the total space, fiber and base of the fibration $\IX$.}
\end{center}
\end{table}

\subsection{A short summary of covariant derivation}\label{sec:derivativetab} 
\vskip-10pt
We have need of derivatives that are covariant under the coordinate transformations
\eqref{eq:diffeomorphisms}, this requires a refinement of the derivatives defined in \citeI for which covariance was required only under the simpler transformations $(y,\,x)\to (\tilde{y}(y),\,\tilde{x}(x))$.
We are led to construct outer derivatives that descend from $\Id$ and covariant derivatives $\Dethsharp_a$ and 
$\fD_a$. For complex manifolds $X$ and $M$ the operators $\Deth$ and $\Dethsharp$ split further into 
$\deth+\dethb$ and $\dethsharp + \dethsharpb$, which are the analogues of the familiar split 
$\dd{\;=\;}\del + \delb$. 

Furthermore, we overload the derivative symbol so that $\Dethsharp_a$, say, should also be covariant with respect to gauge transformations. When we take into account the complex structure of $X$ and $M$, the 
$\Dethsharp_a$ decomposes further into $\fD_\a$ and $\fD_\bb$, which are suitable generalisations of the holotypical derivatives of \citeI. From \sref{subsec:dcomega} we write $\fD$ in place of 
$\Dethsharp$ even when acting on `gauge neutral' objects since no ambiguity arises, and this gives cleaner expressions. For example, we understand that $\fD_\a \o {\;=\;} \Dethsharp_\a \o$.
\vskip20pt
\begin{table}[H]
\renewcommand{\baselinestretch}{1.5}
\begin{center}
\setlength{\extrarowheight}{5pt}
\begin{tabular}{|>{$}c<{$}|>{$}c<{$}|c|}
\hline
\multicolumn{2}{|c|}{\text{Relation}} & Reference\\[3pt] 
\hline\hline
\multicolumn{2}{|c|}{$\Id \;=\; \Deth + \Dethsharp$}& See \SS2.3\\[3pt] 
\hline
&\Dethsharp \;=\; \dd y^a \Dethsharp_a & See \SS2.3\\[3pt]
\hline
\makebox[4cm][c]{$\Deth\;=\; \deth + \dethb$} 
&\makebox[4cm][c]{$\Dethsharp \;=\; \dethsharp + \dethsharpb $} & See \SS2.5\\[3pt] 
\hline
\multicolumn{2}{|c|}{\hskip8pt
\parbox{6.5cm}{\vskip5pt\centering$\fD_a A\;=\; e_a(A) - (\Deth c_a{}^m)A_m - \Deth_A A_a^\sharp$,\\[3pt]
where $\Deth_A A_a^\sharp \;=\; \Deth A_a^\sharp + [A,\, A_a^\sharp]$\vskip5pt}} 
&\parbox[c]{4cm}{\centering See \SS3.1\\ especially eq~\eqref{eq:MixedPart}} \\[3pt]  
\hline
\makebox[4cm][c]{$\Inabla \del_P  =  \, \IGamma{}^Q{}_P \,\del_Q$} 
&\makebox[4cm][c]{$\Inabla e_P = \, \ITheta{}^Q{}_P e_Q$} & See \SS4\\[3pt] 
\hline
\multicolumn{2}{|c|}{$\Inabla^{\LC}$}& Levi--Civita connection \SS4.2\\[3pt] 
\hline
\multicolumn{2}{|c|}{$\Inabla^{\Bi} $}& Bismut connection \SS4.2\\[3pt] 
\hline
\multicolumn{2}{|c|}{$\Inabla^{\Hu} $}& Hull connection \SS4.2\\[3pt] 
\hline
\multicolumn{2}{|c|}{$\fD_\a \Th^\Hu = \nabla\,(\D_{\a})+i\,\nabla\,\fD_\a\o^{(1,1)} $}& Variation of $\Th$. See \SS4.3\\[3pt] 
\hline
\end{tabular}
\capt{5.0in}{tab:derivs}{A table of derivatives used in the paper.}
\end{center}
\end{table}
\newpage
\section{The fibration $\IX$}\label{sec:fibrationccX}
\vskip-10pt
 Our goal is to realise the embedding of heterotic structures $\Het$, corresponding to the tuple $([X,\o,\O], [\ccE,A], [\ccT_X,\Th], H)$,  and their parameter space $M$ inside a single object, which is a fibration $\ccU$. In this paper we will not discuss singularities of the heterotic structure, leaving these global issues for future work.   Differential calculus on $\ccU$ will then account for the variations of a heterotic structure as one moves across $M$. We start by first considering variations of the manifold $X$ over $M$. We will include the vector bundle $\ccE$, three-form $H$ and the constraints on these objects deriving from the anomaly and supersymmetry later. To describe these variations we consider a fibration $\IX$ which represents the family of complex manifolds $X$ with $c_1=0$ and conformally balanced metric over the moduli space.  

\subsection{Defining  $\IX$}
\vskip-10pt
The fibration $\IX$ is pictured as
\beq\label{eq:BundleDef}
 \begin{tikzcd}
  X \arrow[r]  & \IX\arrow[d]\\
                    & M
\end{tikzcd}~,
\eeq
it has the following properties:
\begin{enumerate}
 \item The fibres of $\IX$ corresponding to the manifolds $X$, and all the fibres are diffeomorphic.
 \item There is a orthogonal decomposition of $\ccT_\IX = V_\IX{\,\oplus\,}H_\IX$ where $V_\IX$ corresponds to $\ccT_X$ and $H_\IX$ is an orthogonal complement; sections of $H_\IX$ will correspond to deformations in an appropriate way. 
\end{enumerate}

We introduce local coordinates $u = (y,x)$ for $\IX$, with notation as displayed in the \tref{tab:coords}. This is such that, for a fixed $y$, $V_\IX = \text{span} \{ \del_m \}$ is the subbundle of $\ccT_\IX$ identified with $\ccT_X$.  We also need a notion of a horizontal subbundle $H_\IX$ of $\ccT_\IX$ such  that $\ccT_\IX =  H_\IX \oplus V_\IX$. 
This is facilitated by introducing a projection operator $\p:\ccT_\IX\to \ccT_\IX$:
\beq\label{defcC}
\p\in\mbox{End}(\ccT_{\IX})\simeq\O^{1}(\IX,\ccT_{\IX})~,\quad \p^2~=~\p~, \quad \p|_{V_\IX} = {\rm id}_{V_\IX}.
\eeq
This operator $\p$ is referred to as an Ehresmann connection in the literature.  The most general expression for  projection operator $\p$ satisfying the conditions above is
\beq\label{cCdef}
 \p~=~\dd x^m\otimes\del_m+c_a{}^m\,\dd y^a\otimes\del_m~,
\eeq
for suitable quantities $c_a{}^m$.  This connection provides the desired orthogonal decomposition
\beq\label{eq:subspacesplit}
\ccT_{\IX}~=~H_{\IX}\oplus V_{\IX}~=~\ker{\!\p}\oplus\im{\!\p}~,
\eeq
where the horizontal and vertical subbundles, together with their duals, are spanned by the following basis vectors and forms
\beq\begin{aligned}
 H_{\IX} &\= \mbox{span}\big\{e_a {\=} \del_a - c_a{}^m\,\del_m\big\}~, &
 \quad V_{\IX} &\= \mbox{span}\big\{e_m {\=} \del_m\big\}~,\\[7pt]
 H^*_{\IX} &\= \mbox{span}\big\{e^a {\=} \dd y^a\big\}~, 
 &\quad V^*_{\IX} &\= \mbox{span}\big\{e^m {\=} \dd x^m+c_a{}^m\,\dd y^a\big\}~.
\end{aligned}\label{eq:ebasis}\eeq
We will refer to this as the ``$e$-basis'' for the tangent and cotangent space of $\IX$. This is also referred to as a non--holonomic basis in the literature. This decomposition into horizontal and vertical subspace is referred to as an almost product structure.

The projection operator $\pi$ in the $e$-basis takes the form $\pi {=} e^m \otimes \del_m$. We can define a related tensor $\IL {=} 1{-}2\pi$ which satisfies the property that $\IL^2 {=} 1$. This tensor defines an almost product structure, and is the natural analogue of an almost complex structure. We refer the reader to \cite{YanoBook} for a comprehensive introduction. In the $e$-basis $\IL$ is diagonal with eigenvalues $+1$ for the horizontal subbundle $H_\IX$ and $-1$ for the vertical subbundle $V_\IX$. 

The horizontal subbundle $H_\IX$ is integrable in the sense of Frobenius's theorem if the Lie bracket of two horizontal vectors is also a horizontal vector. That is, 
\beq
[h_1, h_2] \in H_\IX \quad {\rm for} \quad h_1,h_2 \in H_\IX~, 
\label{eq:integrability1}\eeq
A short computation shows this is satisfied if and only if the Lie bracket of the basis vectors $e_a$ vanishes
$$
[e_a,\,e_b] ~=~ 0~.
$$
As we will see, we can view the symbols $c_a{}^m$ as a connection on $M$ with curvature $S$ which is a vector valued two-form with components:
\beq\label{eq:Scurvature}
S_{ab}{}^m  \= [e_a,\, e_b]^m
\= c_a{}^m{}_{,\,b} -  c_b{}^m{}_{,\,a} + c_a{}^n c_b{}^m{}_{,\,n} - c_b{}^n c_a{}^m{}_{,\,n}~.
\eeq
The tensor $\IL$ has a Nijenhuis tensor defined as
\beq
N_\IL \=( \IL^P \del_P \IL^Q - \IL_P{}^Q \Id \IL^P)\del_Q~,\label{eq:nijIL}
\eeq
where $u^P = (y^a,x^m)$ denotes a point in $\IX$ and we have written
$$
\IL^P = \IL_S{}^P \dd u^S~.
$$
Another short computation shows that 
$$
N_\IL = \frac{1}{4}\, S_{ab}\, \dd y^a \dd y^b~.
$$
Hence, vanishing of the Nijenhuis tensor $N_\IL$ is equivalent to vanishing $S_{ab}$ and this is in turn equivalent to the horizontal subbundle being integrable in the sense of \eqref{eq:integrability1}.  Geometrically, it means that $\IX$ is foliated with $H_{\IX}$ being the tangent bundle to codimension six leaves in $\IX$. That is, the tangent space of $M$ is identified with $H_{\IX}$ and $M$ is a submanifold of $\IX$. We will see later that the vanishing of the Lie bracket above is equivalent to demanding deformations of heterotic structures commute.  Frobenius's theorem  implies that given $N_\IL = 0$ we can find a set of coordinates in which $c_a = 0$. Diffeomorphisms that preserve this are of the form $\wt x \to \wt x'(\wt x)$ and $\wt y \to \wt y'(\wt y)$. 

In the following, in the interest of generality -- possibly with a view to describing non-commuting deformation theory -- we will not assume $S$ vanishes except in the final two sections when we apply it to heterotic geometry.

When no additional structures are present, the automorphism group of $\IX$ consists of bundle diffeomorphisms, that preserve the bundle structure. A key property of such a diffeomorphism $\tau$ is that the following diagram should commute:
\beq\label{bundlediffs}
 \begin{tikzcd}
  \IX \arrow[d] \arrow[r, "\t"]  & \IX\arrow[d]\\
  M \arrow[r, "\t^\sharp"]                 & M
\end{tikzcd}~,
\eeq
where  $\t^\sharp:M\to M$ is the naturally induced map. 
Under such a diffeomorphism the fibre $X_{y}$ is mapped onto the fibre $X_{\,\t^\sharp(y)}$ and there is no intersection of fibres. In other words, at least locally, there is a unique manifold $X_y$ corresponding to each point in the base $M$. In our coordinate system $(y^a, x^m)$ a bundle diffeomorphism acts as
\beq\label{eq:BundleDiff}
 (y^a,x^m)\xrightarrow{\ \t\,}\big(\t^a(y),\t^m(y,x)\big)\quad .
\eeq
We can now see that under $\tau$, the symbols $c_a{}^m$ transform as a connection
\beq\notag
 {}^\t c_a{}^m~=~\left(\frac{\del \t{}^m}{\del x{}^n}\,c_b{}^n-\frac{\del \t{}^m}{\del y{}^b}\right)\frac{\del y{}^b}{\del \t^a}~,
\eeq
This is just the transformation law given in \eqref{eq:ctransf}.
With this transformation law the basis elements $e_a, e_m$ in \eqref{eq:ebasis} are covariant and span invariant subspaces. 

Consider a $\tau$ that is the identity map when restricted to the base manifold $M$:
\beq\label{strgrpccFdiagram}
 \begin{tikzcd}
  \IX \arrow[d] \arrow[r, "\vph"]  & \IX \arrow[d]\\
  M \arrow[r, "\text{id}_{M}"] & M
\end{tikzcd}~.
\eeq
This acts as
\beq\label{strgrpaction}
( y^a,x^m )\xrightarrow{\ \vph\,}\big(y^a,\vph^m(y,x)\big)~.
\eeq
These transformations correspond to the action of the structure group on $\IX$. Since $X$ is a manifold the structure group is $\mbox{Diff}(X)$, which is infinite dimensional and non compact, so the bundle $\IX$ is quite different from a principal bundle or a vector bundle. For example, the principal bundle associated to $\IX$ is not a manifold.

\subsection{Tangibility of forms}\label{sec:tangibility}
\vskip-10pt
An $n$-form $\eta$ on $X$ extends to a form $\Ieta$ on $\IX$. In a coordinate basis this takes the form
\beq
\Ieta\= \frac{1}{n!}\,\eta_{a_1 a_2\cdots a_n}\,\dd y^{a_1\cdots a_n}+
\frac{1}{(n-1)!}\,\Ieta_{a_1\cdots a_{n-1} m}\,\dd y^{a_1\cdots a_{n-1}} \dd x^m + \cdots +
\frac{1}{n!}\,\eta_{m_1\cdots m_n}\,\dd x^{m_1\cdots m_n}~,
\notag\eeq
where $\dd y^{a_1a_2\cdots a_n} = \dd y^{a_1} \dd y^{a_2} \cdots \dd y^{a_n}$ and 
$\dd x^{m_1m_2\cdots m_n} = \dd x^{m_1} \dd x^{m_2} \cdots \dd x^{m_n}$.
This expression does not manifestly respect the symmetries of $\IX$ under bundle diffeomorphisms \eqref{eq:BundleDiff}, and so is not convenient. Instead, we will always decompose forms in the $e$-basis
\beq
\Ieta\= \frac{1}{n!}\,\eta^\sharp{}_{a_1 a_2\cdots a_n}\,\dd y^{a_1\cdots a_n}+
\frac{1}{(n-1)!}\,\Ieta^\sharp{}_{a_1\cdots a_{n-1} m}\,\dd y^{a_1\cdots a_{n-1}} e^m + \cdots +
\frac{1}{n!}\,\eta_{m_1\cdots m_n}\,e^{m_1\cdots n_n}~,
\label{splitpform}\eeq
where we adopt the convention of denoting components of forms in this basis by a $\sharp$, and note that  
$\eta^\sharp{}_{m_1\cdots m_n}{\=}\eta_{m_1\cdots m_n}$. We also order forms such that the $\dd y$'s are written first.

We  divide this decomposition into its \emph{corpus} and \emph{animus}: the corpus being the unextended part, 
$\eta_{m_1\cdots m_n}\, e^{m_1}\cdots e^{m_n}$, and the animus being the remainder, including the mixed 
$\dd y^a e^m$ terms. The components of the corpus are always identified with the components of the original unextended form on the manifold $X$. Since we will sometimes want to break up the forms according to their ranks as form on 
$X$ and $M$ we will make reference to the \emph{tangibility} of the parts. We define the tangibility of each term in (\ref{splitpform}) as the pair $[p,q]$  in which $q$ denotes the number of corporal indices. 
\beq\notag
\Ieta~=~ \sum_{p+q = n}\Ieta^{[p,q]}~.
\eeq 
Thus the corpus of $\Ieta$ has tangibility $[0,n]$ while the animus is the remainder, including the mixed terms:
The key virtue of this decomposition is that different tangibilities do not mix under bundle diffeomorphisms. 
 We use square brackets to distinguish tangibility from holomorphic type. 

\subsection{Differential calculus on $\IX$}
\vskip-10pt
We now introduce a differential calculus on $\IX$ that will eventually describe how heterotic structures vary over the moduli space $M$. 

The de Rham operator on $\IX$, which defines the exterior derivative, is denoted by
\beq\notag
 \Id:\O^{n}(\IX)\to\O^{n+1}(\IX)~.
\eeq
When acting on the basis of forms $\{\dd y^a,e^m\}$ it gives\footnote{We are redefining the operator $\dd$, so the differentials, hitherto denoted by $\dd y^a$, should now be written~$\Id y^a$. Since $\dd$ now denotes the $x$-part of $\Id$, we have $\dd(y^a){\;=\;}0$. We will however abuse notation by continuing to write 
$\dd y^a$, when we mean $\Id y^a$.}
\beq\notag
\Id(\dd y^a)\= 0~,\quad \Id e^m\=
- \del_n c_a{}^m\,\dd y^a e^n-\frac{1}{2}S{}_{ab}{}^m\,\dd y^a \dd y^b~,
\eeq
where the second relation has parts of tangibility $[1,1]$ and $[2,0]$. This implies that $\Id$ acting on a $[0,q]$ form $\eta$ is
\beq\notag
 \Id\eta=(\Id\eta)^{[0,q+1]}+(\Id\eta)^{[1,q]}+(\Id\eta)^{[2,q-1]}~,
\eeq
where the different tangibilities are given by
\beq\label{actionId}\begin{split}
(\Id\eta)^{[0,q+1]}&\= 
\frac{1}{q!}\,(\del_{n}\eta_{m_1\cdots m_q})\,e^{n} e^{m_1\cdots m_q}~,\\[10pt]
(\Id\eta)^{[1,q]~~~}&\= \frac{1}{q!}\Big(e_a(\eta_{m_1\cdots m_q}) - 
c_a{}^n{}_{,m_1}\eta_{n m_2\cdots m_q} - \cdots 
- c_a{}^n{}_{,m_q}\eta_{m_1\cdots{m_{q-1}}n}\Big)\,\dd y^a e^{m_1\cdots m_q}~,\\[10pt]
(\Id\eta)^{[2,q-1]}&\= -\frac{1}{2(q-1)!}\,S{}_{ab}{}^{n}\,\eta_{nm_1\cdots m_{q-1}}\,\dd y^{ab} e^{m_1\cdots m_{q-1}}~.
\end{split}\raisetag{24pt}\eeq
where in the second line we have denoted the non-holonomic derivative as $e_a(\eta_{m_1\cdots m_q}) {=} (\del_a- c_a{}^n \del_n)\eta_{m_1\cdots m_q}$. This is also referred to as a pfaffian derivative in the mathematics literature. 
We see that, in general, a tangibility $[p,q]$ form, with $q\ge 1$,  is mapped to a form with three different tangibilities divided up into branches as shown\beq\notag
\begin{tikzcd}[row sep=2em, column sep=3em]
&\O^{[p, q+1]}(\IX) \\
\Id :~\O^{[p,q]}(\IX) \arrow[ru, "\Deth", xshift=2.5ex] \arrow[r, "\Dethsharp"] \arrow[rd, "S", xshift=2.5ex] 
&\O^{[p+1, q]}(\IX) \\
&\O^{[p+2, q-1]}(\IX) \\
\end{tikzcd}~.
\eeq
\vskip-30pt
The presence of three possible branches  is analogous to a non-integrable complex structure in which the exterior derivative of a holomorphic type $(p,q)$ form has non-vanishing $(p+2,q-1)$ and $(p-1,q+2)$ projections which depend on the Nijenhuis tensor. Note that, we get a reduction to two tangibilities as the horizontal subbundle is integrable but we have illustrated the presence of $S$ for generality. 

The first two branches are defined by two covariant derivatives $\Deth$ and $\Dethsharp$ 
\beq\label{eq:ExteriorDef}
\begin{split}
\Deth\ &:\O^{[p,q]}(\IX)\to\O^{[p,q+1]}(\IX)~,\\[0.2cm]
\Dethsharp &:\O^{[p,q]}(\IX)\to\O^{[p+1,q]}(\IX)~.
\end{split}
\eeq
They are defined by  specifying their action on a zero-form $f$, a function, and on the basis
$\{\dd y^a,e^m\}$ of one-forms\footnote{Note that it is not the case that $\Deth{\;=\;} e^m\del_m$ and 
$\Dethsharp{\;=\;}\dd y^a \del_a$.}:
\beq\begin{aligned}
\Deth\, f &\= (\del_m f)\,e^m~,\qquad & \Deth\, e^m &\= 0~,\quad & \Deth\,(\dd y^a) &\= 0~,\\[5pt]
\Dethsharp f &\= e_a(f)\,\dd y^a~,\qquad & \Dethsharp e^m &\= - c_a{}^m{}_{,\,n}\,\dd y^a e^n~,
\hskip10pt & \Dethsharp(\dd y^a) &\= 0 ~.
\end{aligned}\label{eq:dSharp}\eeq
It is straightforward to check that
\beq\notag
\Deth^2 \= 0~~~\text{and}~~~\{\Deth,\,\Dethsharp\}\= 0~.
\eeq
It is straightforward to check that the second line of \eqref{actionId} is consistent with applying the rules \eqref{eq:dSharp} to compute $\Dethsharp\eta$: 
\beq\notag
\Dethsharp\eta\= \dd y^a\,\Dethsharp_a\eta~,
\eeq
where we have defined a shorthand  
\beq\begin{split}
\Dethsharp_a\eta &\= \frac{1}{q!}\Big(e_a(\eta_{m_1\cdots m_q}) - 
c_a{}^n{}_{,\,m_1}\,\eta_{n m_2\cdots m_q} - c_a{}^n{}_{,\,m_2}\,\eta_{m_1 n\cdots m_q} - \cdots\Big)\,e^{m_1}\cdots e^{m_q} \\[5pt]
&\= e_a(\eta) - (\Deth c_a{}^m)\,\eta_m~.
\end{split}\label{eq:delsharpeta}\eeq
In passing to the second line we have used $e_a \cdot e^m = 0$ and 
\beq
\eta_m\= \frac{1}{(q-1)!}\,\eta_{mn_1...n_{q-1}} e^{n_1} \cdots e^{n_{q-1}}~.
\notag\eeq
The derivative $\Dethsharp_a \eta$  describes a variation of $\eta$ in the direction of a vector $e_a$, along the moduli space. Notice that this is identified with the Lie derivative $\cL_{e_a} \eta$. 

Both the $\Deth$-operator and $\Dethsharp$-operator are invariant under diffeomorphisms. This is most easily seen by projecting onto tangibility, we have for example $\Dethsharp\eta{\=}(\Id\eta)^{[1,q]}$.

In \citeI we implicitly assumed that variations could be written by partial derivatives, for example that the first order deformation of $\o$ is given by \hbox{$\d \o {\=} \dd y_a \del_a\o$}. 
Now we see that we should write
\beq
\d\o \= \dd y^a\Dethsharp_a \o~.
\notag\eeq
We also have need for second order variations of fields. In this case, the curvature $S$ introduces a commutator:
$$
[\Dethsharp_a, \Dethsharp_b ] \o \= S_{ab}{}^m (\Deth \o)_m + \Deth (S_{ab}{}^m\, \o_m)~. 
$$
So we see that demanding that $\p$ be integrable, so that $S=0$, is equivalent to demanding that all variations commute and that $(\Dethsharp)^2 {\=} 0$. The bundle then becomes locally flat. As previously mentioned we will keep $S$ around in the interest of generality, and set it to zero at the end.

To summarise: in general, $\Id\eta$ decomposes into three terms,
\beq
 \Id\Ieta\= \Deth\Ieta + \Dethsharp\Ieta - 
 \frac{1}{2 q!}\,\dd y^{a_1}\cdots \dd y^{a_{q-1}} \dd y^b \dd y^c\,S{}_{bc}{}^n\,\Ieta^\sharp_{a_1\cdots a_{q-1}\,n}~.
\label{eq:Idsplit}\eeq
with the parts given by \eqref{eq:Idsplit} as above. We have  the identities
\beq
\Id^2 \= 0~,~~ \Deth^2 \= 0~~~\text{and}~~~\{\Deth,\,\Dethsharp\}\= 0~.
\notag\eeq 
If, in addition, the curvature $S$ vanishes, then we have also 
\beq
\Id \= \Deth + \Dethsharp~~~\text{and}~~~(\Dethsharp)^2 \= 0~.
\notag\eeq

\subsection{Derivatives of tensors}
\vskip-10pt
Consider a tensor $\xi$ whose indices are purely vertical:
\beq\notag
 \x \= \x{}_{n_1 \cdots  n_s}{}^{m_1 \cdots  m_r}\ e^{n_1}\otimes\cdots \otimes e^{n_s}\otimes\del_{m_1}\!\otimes\cdots \otimes\del_{m_r}~.
\eeq
The deformation of this tensor in the direction of a horizontal vector $Y^\sharp = Y^a(y)\, e_a$ is given by a Lie derivative 
\beq\notag
 \cL_{Y^\sharp}\xi \= Y^a \cL_{e_a}\xi~.
\eeq
We take horizontal vector fields $Y^\sharp$ to depend only on the parameters. This Lie derivative then acts as a directional derivative along the moduli space.\footnote{This does not hold generally for the Lie derivative. For instance the $\ccL_{fY}(X) = f \ccL_{Y}(X) - Y(f) X$, where $f$ is a function and $X,Y$ two arbitrary vectors.}

It is useful to describe the action of the Lie derivative in this way in components. Its action in the direction $e_a$  is determined by
\beq\label{eq:cLonBases0}
 \cL_{e_a} f \= e_a (f)~, \quad \cL_{e_a}(\del_m)\= c_a{}^n{}_{,\,m}\,\del_n~,\quad 
 \cL_{e_a}(e^m)\= - c_a{}^m{}_{,\,n}\,e^n - S_{ab}{}^m \dd y^b~.
\eeq
Therefore, the Lie derivative of $\x$ along the horizontal vector $e_a$ has vertical components given by
\beq\label{varxicomponents}
\begin{split}
\cL_{e_a}\,\x{}_{n_1 \cdots  n_s}{}^{m_1 \cdots  m_r}
&\= e_a(\x{}_{n_1 \cdots  n_s}{}^{m_1 \cdots  m_r})+
c_a{}^{m_1}{}_{,\,k}\, \x{}_{n_1 \cdots  n_s}{}^{k m_2\cdots  m_r} + \cdots +
c_a{}^{m_r}{}_{,\,k}\, \x_{n_1 \cdots  n_s}{}^{m_1 \cdots  m_{r-1}k} \\[0.1cm]
&\hskip30pt - c_a{}^k{}_{,\,n_1}\, \x{}_{k n_2 \cdots  n_s}{}^{m_1 \cdots  m_r} - \cdots -
c_a{}^k{}_{,\, n_s}\, \x{}_{n_1 \cdots  n_{s-1} k}{}^{m_1 \cdots  m_r}~.
\end{split}\raisetag{18pt}
\eeq
When applied to a differential form $\eta$ reduces to the form given by \eqref{eq:delsharpeta}. 

We can relate this to the derivative $\Dethsharp$. The covariant derivative with respect parameters is 
\beq\label{eq:cLonBases}
\Dethsharp_a f\= e_a(f)~,\quad \Dethsharp_a\,
\del_m\= c_a{}^n{}_{,\, m}\,\del_n~,\quad
\Dethsharp_a\, e^m\= - c_a{}^m{}_{,\, n}\,e^n~,
\eeq
and this is  the projection of \eqref{eq:cLonBases0} onto the fibers. This amounts to dropping the term proportional to $S$ so, when $\p$ is integrable, derivatives of tensors are described exactly~by~$\Dethsharp$.

\subsection{The Fr\"olicher--Nijenhuis bracket}
\label{s:FN}
\vskip-10pt
The Fr\"olicher--Nijenhuis (FN) bracket extends the Lie bracket on vector fields to vector field valued forms:
\beq\notag
 [~,\ ]_{FN}:\O^k(\IX,\ccT_{\IX})\times\O^l(\IX,\ccT_{\IX})\to\O^{k+l}(\IX,\ccT_{\IX})~,
\eeq
We are interested in the case $k{\=}l{\=}1$. Given $\IK,\IL \in \O^1(\IX, \ccT_\IX)$ and vectors 
        $\Iv,\Iw$ we have
\beq\notag
 \begin{split}
  [\IL,\IK]_{FN}(\Iv,\Iw) &{=} [\IL \Iv, \IK\Iw] {+} [\IK \Iv, \IL\Iw] {+} (\IK\IL + \IL\IK)[\Iv,\Iw] {-} \IK([\IL\Iv,\Iw] {+} [\Iv,\IL\Iw]) {-} \IL([\IK\Iv,\Iw] {+} [\Iv,\IK\Iw])\,,
 \end{split}
\eeq
where the brackets on the right-hand side are the usual Lie brackets. 
The  Nijenhuis tensors of $\IL$ and $\IK$ are expressible in terms of the Fr\"olicher-Nijenhuis bracket:
\beq\notag
 \begin{split}
 N_\IL\= \frac{1}{2}\,[\IL,\IL]_{FN}~, \qquad  N_\IK\= \frac{1}{2}\,[\IK,\IK]_{FN}~.\\
 \end{split}
\eeq

\subsection{The metric and complex structure on $\IX$}\label{s:metricCS}
\vskip-10pt
We now put a metric and complex structure on $\IX$ in a way that appropriately reflects the fact we are studying the moduli of heterotic structures. 
We start with a metric on $\IX$. For the vertical fibers $X_y$ we have a family of metrics 
$g_{mn}(y,x)\dd x^m\otimes\dd x^n$, and for the moduli space $M$ we have a natural metric 
$g^\sharp_{ab}(y)\ \dd y^a\otimes\dd y^b$, which, correct through $\cO(\ap)$, is given by
\eqref{eq:modulimetric}. The most straightforward way to combine these parts is to require the metric 
to be compatible with the projection operator $\p$ which amounts to the metric being block diagonal in the $e$-basis. 
\beq
\dd s^2 \= g_{mn}(y,x)\ e^m\otimes e^n + g^\sharp_{ab}(y)\ \dd y^a\otimes\dd y^b~.
\label{eq:ccFMetric}\eeq
With respect to this metric, the spaces $V_\IX$ and $H_\IX$ are orthogonal.

In more erudite language: we note that since $V_{\IX}\cong\ccT_{X}$ we should identify the inner product on $V_{\IX}$ with the metric on $X$. For the horizontal part, first note that $\ccT_{M}|_{y}$ and 
$H_{\IX}|_{(y,x)}$ are isomorphic as vector spaces. Demanding compatibility between the two projections
$\ccT_{\IX}\to\ccT_{M}$ and $\ccT_{\IX}\to H_{\IX}$~, so the horizontal metric needs to coincide with the moduli metric. This leads to the form given.

The given form of the metric is not unique, since the components of the connection $c_a{}^m$ are not specified. In a coordinate basis we have
\beq
\Ig_{ma} \= g_{mn} c_a{}^n~.
\notag\eeq
So the off-diagonal blocks of the metric have been left, so far, unspecified.

The fibers of $\IX$ and also the base $M$ are complex manifolds; we ask if $\IX$ is also. Denoting the complex structures of $X$ and $M$ by $J$ and $J^\sharp$, we can combine them into an almost complex structure $\IJ$  for $\IX$ in a manner analogous to the construction of the metric
\beq\label{IJ}
 \IJ\= J^\sharp{}_b{}^a\ \dd y^b\otimes e_a+J_n{}^m\ e^n \otimes \del_m~,
\eeq
This is tantamount to demanding that $\IJ$ and $\p$ commute as endomorphisms, but not as differential structures. Demanding they commute as differential structures means, essentially, that $[\IJ,\IL]_{FN} = 0$, which is too strong for our situation. It is immediate that $\IJ^2{\;=}-\one$, without further conditions on the connection $c_a{}^m$.

We examine next the Nijenhuis tensor of $\IJ$. In Appendix \ref{app:Nij} we show that this has the form
\beq\notag
N_\IJ \; =\, -2\Big( 2 [P^\sharp_a, P_m]^q \,e^a \,e^m\, Q_q + 2[Q^\sharp_a, Q_m]^q \,e^a\, e^m\, P_q 
+[P^\sharp_c, P^\sharp_d]^q\, e^c \,e^d\, Q_q + [Q^\sharp_c, Q^\sharp_d]^q \,e^c \,e^d \,P_q \Big)~,
\eeq
where $P_m = P_m{}^n \del_n$ is a vector constructed out of the projector $P_m{}^n$, and $P^\sharp_a$ is constructed analogously.  The term $[P^\sharp_c, P^\sharp_d]$ is the $(2,0)$-component of the curvature $S$.  
 If the manifold is complex, which is the case $N_\IJ{\=}0$, then we can write these terms in complex coordinates  $(y^\a,y^\bb,x^\m, x^\nb)$.  The vanishing of the right hand side requires
\beq
\quad [e_\a, \del_\m]^\nb \= 0~, \quad\text{and}\quad [e_\a, e_\b]^\nb \= 0~.
\notag\eeq
These equations are then constraints on the connection $c_a{}^m$:
\beq\begin{split}
0&\= [e_\a,e_\m]^\nb \;= - \del_\m c_\a{}^\nb ~,\\[5pt]
0&\= [e_\a, e_\b]^\nb \= c_\a{}^\nb{}_{,\,\b} - c_\b{}^\nb{}_{,\,\a} + 
c_\a{}^\rb\, c_\b{}^\nb{}_{,\,\rb} - c_\b{}^\rb\, c_\a{}^\nb{}_{,\,\rb}~,
\label{eq:INeqs}\end{split}\eeq
where, in writing the second equation, we have omitted terms that vanish as a consequence of the first equation.

Furthermore, both $\IJ$ and $\p$ are elements of End$(\ccT_{\IX})$. As their action commutes they can be diagonalised simultaneously, inducing a further split
\beq\notag\begin{split}
\ccT_{\IX}\= H^{(1,0)}_{\IX}\oplus H^{(0,1)}_{\IX}\oplus V_{\IX}^{(1,0)}\oplus V^{(0,1)}_{\IX}~,
\end{split}\eeq
which is realised through
\beq\notag
H^{(1,0)}_{\IX}\= \mbox{span}\big\{e_\a =\,\del_\a-c_\a{}^\m\,\del_\m\big\}~,~~~
V^{(1,0)}_{\IX}\= \mbox{span}\big\{\del_\m\big\}~,
\eeq
together with their complex conjugates. In fact, in the first line one could have included a mixed term 
$-c_\a{}^{\nb}\,\del_{\nb}$. This vanishes  because  the following relation needs to hold if $\IX$ is complex
\beq\notag
 \mbox{span}\big\{e_\a,\,\del_\m\big\}\= \mbox{span}\big\{\del_\a,\,\del_\m\big\}~.
\eeq
Note that if
\beq
c_\a{}^\nb \= 0~,
\notag\eeq
then equations \eqref{eq:INeqs} are both satisfied, without further conditions on the $c_a{}^m$.

We can decompose $\Deth$ and $\Dethsharp$ into holomorphic type
\beq\label{eq:sharpenedDel}\begin{split}
\Deth~         &\= \deth \,+\, \dethb~,\\[5pt]
\Dethsharp   &\= \dethsharp + \dethsharpb \=  \dd y^\a\,\Dethsharp_\a + \dd y^{\bb}\,\Dethsharpb_{\bb}~,
\end{split}\eeq
where $\deth,\dethb$ both square to zero and anticommute, and are Dolbeault operators with respect to $J$. That is, $\dethb$ maps a $J$--$(p,q)$ form to a $J$--$(p,q+1)$-form. Their close relatives, $\dethsharp$ and 
$\dethsharpb$  anticommute when $S_{\a\bb}{}^m {\=} 0$ and $\dethsharpb$ squares to zero when 
$S_{\ab\bb}{}^m{\=}0$. When this is the case $\dethsharpb$ is a Dolbeault operator with respect to $J^\sharp$ mapping a $J^\sharp$--$(p,q)$ form to a $J^\sharp$--$(p+1,q)$~form. 

Recall that in special geometry and Kodaira--Spencer theory, the parameter variation of complex structure and the integrability condition $N_J = 0$, to first order, becomes
\beq\label{eq:DCohomology}
 \D_\a{}^\m\in H^{(0,1)}_{\delb}(X,\ccT^{(1,0)}_{X})~.
\eeq
The covariant deformation of the holomorphic projector $P_m{}^n {=} \half (\d_m{}^n {-} \ii J_m{}^n)$ is given by the Lie derivative \eqref{varxicomponents}, and this facilitates a covariant definition of $\D_\a$:
\beq\label{eq:DefDelta}
  \D_a \= \cL_{e_a} P \= \Big(e_a(P_n{}^m)+
 (\del_k c_a{}^m)P_n{}^k-(\del_n c_a{}^k) P_{k}{}^m\Big) e^n \otimes \del_m~.
\eeq
If $N_\IL = 0$ so that the product structure is integrable, then we can find  a set of adapted coordinates in which $c_a{}^m = 0$ and that $ \D_a = \del_a P_m{}^n \dd x^m \del_n$. This is what is familiar from special geometry. Provided $\del_a J_m{}^n \ne 0$, so that complex structure depends on parameters, we cannot find a set of holomorphic coordinates in which $c_a$ is also zero.

On the other hand, as $N_\IJ = 0$ we can find a set of holomorphic coordinates in which $\IJ$ is constant and diagonal. In that case 
\beq
\cL_{e_\a} P_\nb{}^\m \=\!- \del_\nb\, c_\a{}^\m~. 
\eeq
 In this coordinate chart we identify
 \beq\label{Deltaisdelc}
 -\dethb c_{\a}{}~=~ \D_{\a}{}^{\m}\del_\m~.
\eeq
 Although the symbols $c_\a{}^{\m}$ transform in the manner of a connection, the quantity $\del_\nb c_\a{}^\m$ is covariant under holomorphisms and so is a well-defined $\dethb$-closed form, consistent with \eqref{eq:DCohomology}. It is important to note that we cannot set  $c_\a{}^\m= 0$ since $\D_\a$ is a tensor which does not vanish. 
 
 If we have $N_\IL = N_\IJ = 0$ then this does not necessarily imply we can find a set of holomorphic coordinates for $\IX$ in which $c$ vanishes. Indeed, as we see from the above $\cL_a J = [e_a, J]_{FN}$ is precisely the obstruction to doing this. Instead, the structures $\IJ$ and $\IL$ are simultaneously integrable if 
 $$
 [\IJ, \IJ]_{FN} \=0~, \quad [\IL,\IL]_{FN} \= 0~, \quad [\IJ,\IL]_{FN} \= 0~.
 $$
While we will often use the first two conditions, we will always have $[\IJ,\IL]_{FN}$ being non-vanishing and so we cannot discard $c$. 
 
 In \cite{Candelas:1987se,Candelas:1990pi} it is observed that given $\{e_\a\}$ define normal vectors to fibres $c_\a{}^\m$ with respect to the metric on $\IX$. Hence, $c_\a{}^\m$ is related to the extrinsic curvature  by $(\del_\mb c_\a{}^\r) g_{\r\nb}$ which describes the curvature of fibres are embedded within $\IX$. It is also shown that deformations of complex structure satisfy exactly the relation \eqref{Deltaisdelc}. We discuss this further below.

The $\IJ$-Dolbeault operator $\Idel$ acting on a  form of $\IJ$-type $(p,q)$, denoted $\IW^{(p,q)}$,  is defined as 
 $$
 \Idel\, \IW^{(p,q)} \= (\Id \IW)^{(p+1,q)}~.
 $$  
In the language of $\IX$, $\del_\a$ is promoted to a Lie derivative  $\cL_{e_\a}$ while holomorphic type of vertical forms are defined by the  projectors 
\beq\label{eq:projectors}
P \= \half( \d_m{}^n - \ii J_m{}^n)\,e^m \otimes \del_n~, \quad \text{and} \quad 
Q \= \half( \d_m{}^n + \ii J_m{}^n)\,e^m \otimes \del_n~.
\eeq
We have $\cL_{e_\a} P = \D_\a$ and $\cL_{e_\a} Q = - \D_\a$. On a corporal 1-form $\eta = \eta_m e^m$, 
the $\Dethsharp_\a$ operator
is the appropriate projection of the Lie derivative: 
\beq
\Dethsharp_\a \eta^{(p,q)} \= (\cL_{e_\a} \eta)^{(p,q)}~. 
\notag\eeq
This is precisely the holotypical derivative denoted by $\ccD_\a$ in \citeI. 
As may be seen by writing out the components for a 1-form, for example
 \beq\label{eq:holotypicalDef}
 \begin{split}
 \Dethsharp_\a\eta^{(1,0)} ~&=~ (\cL_{e_\a} \eta_m)\, P_n{}^m e^n \,\=  \cL_{e_\a} (P_n{}^m \eta_m e^n) - \D_\a{}^m \eta_m~,\\[5pt]
 \Dethsharp_\a\eta^{(0,1)} ~&=~  (\cL_{e_\a} \eta_m)\, Q_n{}^m e^n \= \cL_{e_\a} (Q_n{}^m \eta_m e^n) + \D_\a{}^m \eta_m ~.
  \end{split}
\eeq
The second equality follows by the Leibniz rule.  

\subsection{The connection $c_a{}^m$ as the shift,  and the extrinsic curvature of~$X_y$}\label{s:xtrinsic}
\vskip-10pt
Recall first the formalism relating to the extrinsic curvature of a submanifold as it applies in an elementary setting such as in \fref{fig:xtrinsicsimple}. Here we have a curve which we approximate to second order by a circle of curvature. At a point on the curve we have a normal $\d{\bf n}{\=}\d r\, {\bf n}$ 
(${\bf n}{\=}\del/\del r$)
 and a tangent ${\bf m}\,({=\;} \del/\del\th)$. We consider the result of parallely propagating $\d\bf n$, in the embedding metric, to a nearby point, this gives the dashed vector in the figure. The pre-existing normal $\d\bf n$ at the displaced point differs from this by an amount proportional to $\bf m$. We write\footnote{There is a choice of sign here. Our choice makes the extrinsic curvature of a cylinder in a flat embedding space positive. An opposite convention is also common.}
\beq
{\bf m}\cdot\!\Inabla{\bf n}\= \chi\, {\bf m}~,
\notag\eeq
and this defines the extrinsic curvature $\chi$. Either from the diagram, or from a direct calculation of the covariant derivative, one sees that $\chi{\=}1/r$ for the situation depicted.

We could study also the variation of the tangent vector, by parallely propagating the tangent vector to a nearby point on the surface and comparing it with the preexisting tangent vector there. In this way we see that 
\beq
{\bf m}\cdot\!\Inabla{\bf m}\;=\, - \chi\, {\bf n}~.
\notag\eeq
Again, we can check this directly by computing the covariant derivative, and we can also deduce this relation by noting that the right hand side is in the direction of $-{\bf n}$ and the coefficient follows from the previous relation, on noting that ${\bf m}\cdot\!\Inabla{(\bf m\!\cdot \!n)}{\=}0$.

We turn now to \fref{fig:prextrinsic}, which relates to the fibration $\IX$. We first consider the case that the quantity $c_a{}^m$ vanishes, so that the normals connect the points labeled by $x$ on $X_y$ and 
$X_{y+\d y}$ and also the points labelled by $x+\d x$. We may parallely propagate the normal $e_a$ from 
$x$ to $x+\d x$, on $X_y$, and compare it with the preexisting normal there. In this way we can define and extrinsic curvature tensor $\chi_{am}{}^n$
\beq
e_m\cdot\Inabla e_a \= \chi_{am}{}^n\, e_n~.
\label{eq:xtrinsicNormal}\eeq
We can study the variation of the tangents rather than the variation of the normals. We take a tangent vector $e_n$ at $x$, parallely propagate it to $x+\d x$ and compare it with the pre-existing tangents. There will be an out of surface component that can be expressed in terms of the normal vectors. This process yields
\beq
e_m\cdot\Inabla e_n \;=\, -\chi^a{}_{mn}\, e_a + \IGamma_m{}^k{}_n\, e_k~.
\notag\eeq
The fact that the coefficients that involve the $e_a$, on the right hand side, are the extrinsic curvature follows from \eqref{eq:xtrinsicNormal}, on noting that $e_m\!\cdot\!\Inabla \Ig(e_a,\,e_n){\=}0$. We see also, in this way, that the $a$ index on the extrinsic curvature is raised and lowered with the metric $g_{ab}$, while the $n$ index is raised and lowered with the metric $g_{mn}$.

The extrinsic curvature so defined is a tensor, so covariant and so unaffected by whether we choose to take $c_a{}^m{\=}0$. However, if we do so, we can identify the extrinsic curvature with minus the Christoffel symbol $\IGamma_m{}^n{}_a$.

Now let us include the effect of nonzero $c_a{}^m$ and turn to \fref{fig:xtrinsic}. The vector $\d y^\a \del_\a$ now connects the two points $(y,\, x)$ and $(y+\d y,\,x)$. So the point labeled by $x$ on $X_y$ with the point labeled by $x$ on $X_{y+\d y}$. The normal vector $\d y^\a e_\a$ connects the point $x$ on $X_y$ with the point $x - c_\a \d y^\a$ on $X_{y+\d y}$. The difference is the vertical vector $c_\a \d y^\a$. Following the usage in relativity, we refer to $c_\a$ as the \emph{shift}. For the displaced point $x+\d x$, the shift has become 
$(c_\a + \d c_\a)\d y^\a$. 
For the case of real coordinates, we have the freedom to take $c_a{}^m{\=}0$. In complex coordinates, however, this is no longer possible. Indeed the shift plays an essential~role. 

We see from \eqref{eq:xtrinsicNormal} that the extrinsic curvature is a rotation coefficient
\beq
 \chi_{am}{}^n \= \ITheta_a{}^n{}_m \= \frac12 g^{n k}\fD_a g_{km}~,
\notag\eeq
where $\ITheta$ are the rotation coefficients in the $e$-basis, and the last term follows from computing the Levi--Civita connection coefficient in the $e$-basis, see Appendix \sref{app:LeviCivita}, with 
\beq
\fD_a g_{km} \= e_a(g_{km}) - c_a{}^\ell{}_{,\,k}\, g_{\ell m} - c_a{}^\ell{}_{,\,m}\, g_{\ell k}~.
\notag\eeq
In complex coordinates, it follows that we have
\beq
\chi_{\a\mb}{}^\n \= \frac12 g^{\n\kb} \fD_\a g_{\kb\mb} \= 
-\frac12\, g^{\n\lb}\Big( c_\a{}^\r{}_{,\,\lb}\, g_{\r\mb} + c_\a{}^\r{}_{,\,\mb}\, g_{\r\lb} \Big)~.
\label{ChiInTermsOfC}\eeq
The last term in this equation expresses the extrinsic curvature in terms of the derivatives of the shift.

We also know from \eqref{Deltaisdelc} that $c_\a{}^\r{}_{,\,\lb} {\;=\,} - \D_{\a\lb}{}^\r$. So we also have the following expression for the extrinsic curvature in terms of $\D_\a{}^\n$
\beq
\chi_{\a\mb\nb} \= \D_{\a\,(\mb\nb)}~. 
\notag\eeq

\vskip20pt
\setlength{\fboxsep}{0.2in}
\setlength{\fboxrule}{1pt}
\begin{figure}[H]
\centering
\vbox{
\framebox{\parbox[c]{6.05in}{\vskip.1in\hskip1in\includegraphics[width=4in]{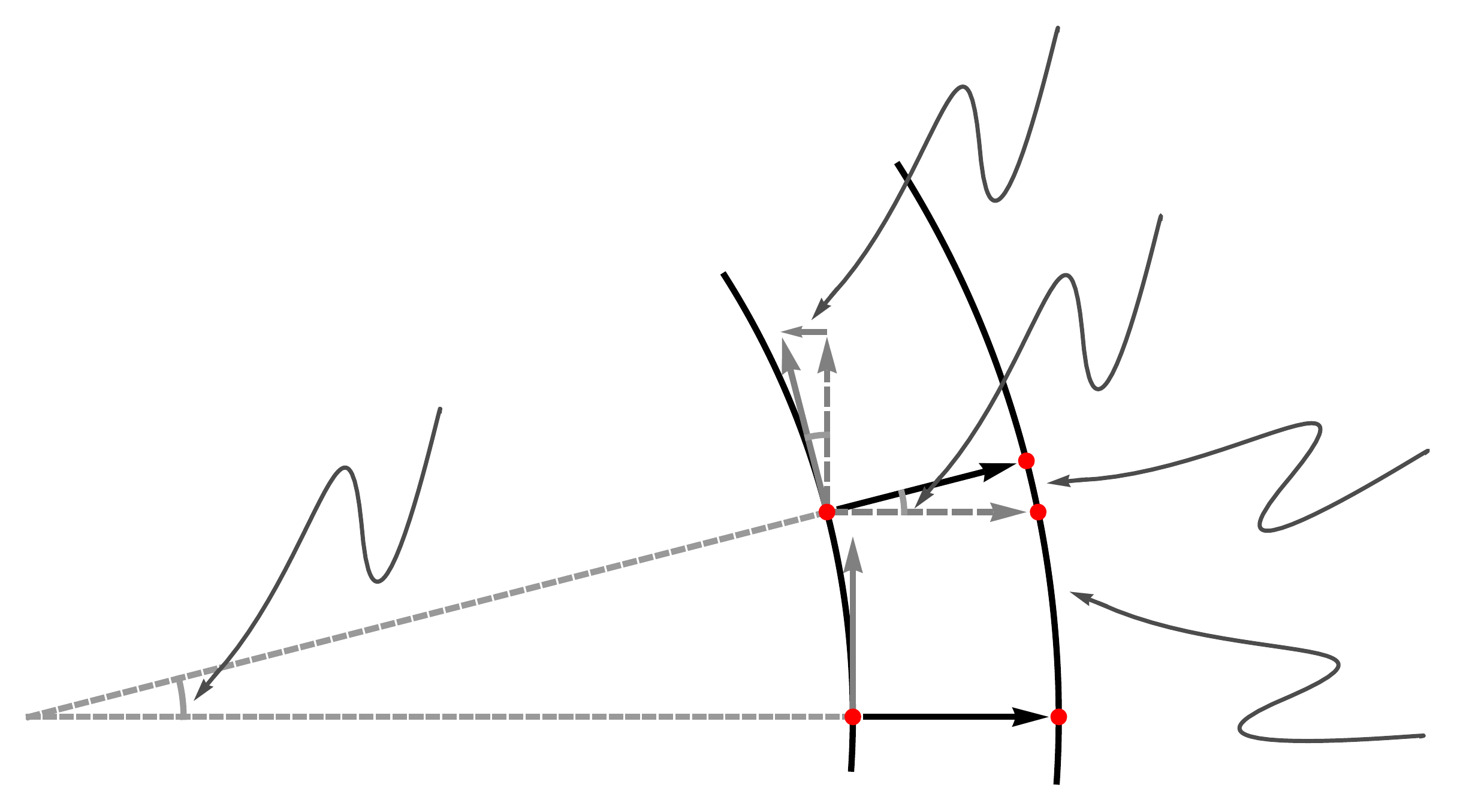}\hskip1in}}
\vskip5pt
\place{2.4}{1.55}{\large$\d\th$}
\place{4.4}{2.05}{\large$\d\th$}
\place{5.2}{1.4}{\large$\d r\d\th$}
\place{5.2}{0.53}{\large$r\d\th$}
\place{2.7}{0.4}{\large$r$}
\place{3.75}{0.4}{\large$\d r$}
\place{3.9}{2.6}{\large$ -\d\th\hskip1pt{\bf n}$}
\capt{6in}{fig:xtrinsicsimple}{The calculation of the extrinsic curvature for a cylinder embedded in a flat space.}
}
\end{figure}
\vfil
\setlength{\fboxsep}{0.5in}
\setlength{\fboxrule}{1pt}
\begin{figure}[H]
\centering
\vbox{
\framebox{\parbox{5.45in}{\vskip-5pt\hskip1.5cm\includegraphics[width=3.09in]{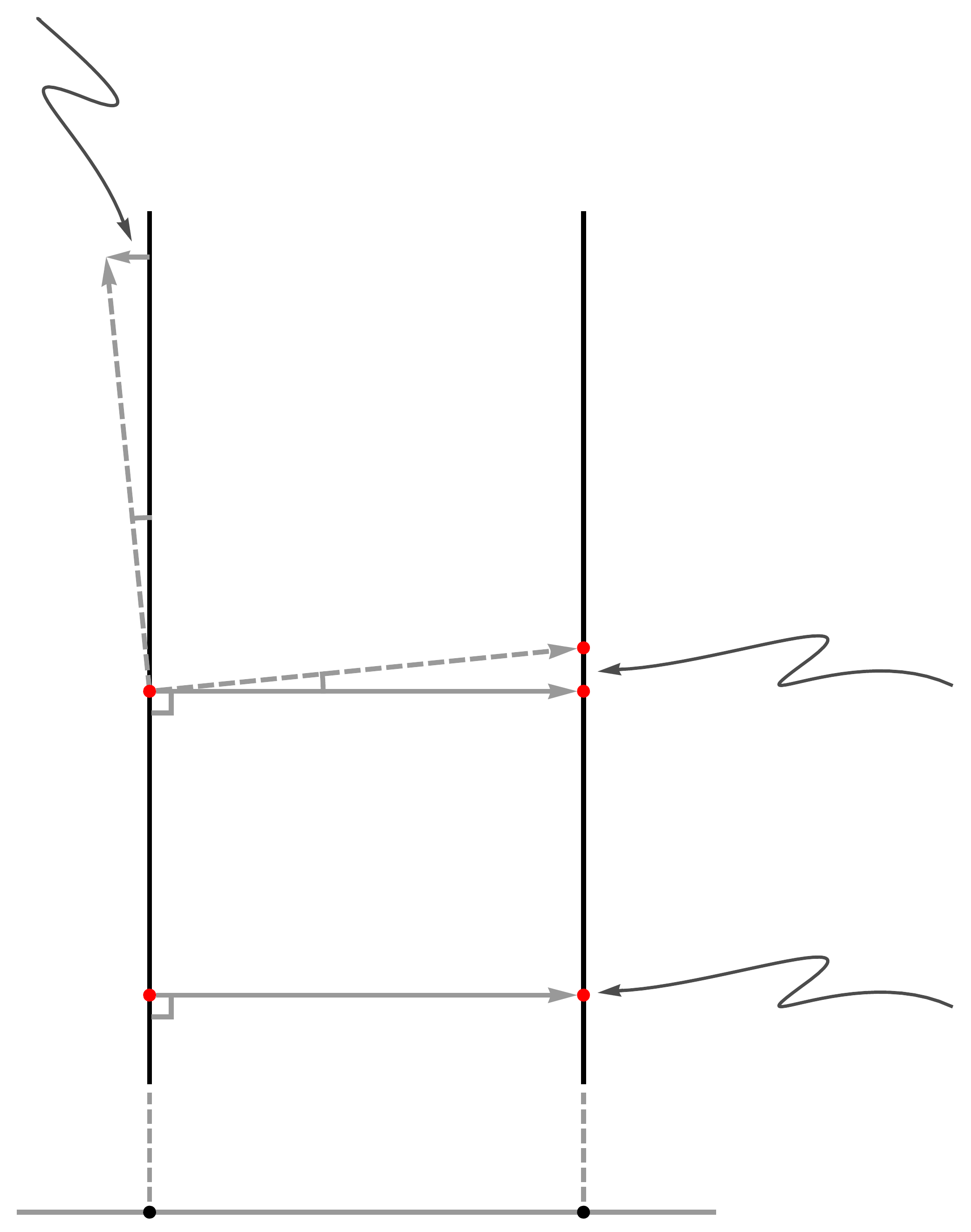}\hskip0.9in\vskip-10pt}}
\vskip29pt
\place{1.38}{1.5}{\large$x$}
\place{0.95}{2.5}{\large$x+\d x$}
\place{1.52}{4.2}{\large$X_y$}
\place{2.9}{4.2}{\large$X_{y+\d y}$}
\place{0.8}{0.85}{\large$M$}
\place{1.52}{0.65}{\large$y$}
\place{2.77}{0.67}{\large$y+\d y$}
\place{2}{1.7}{\large$\d y^a \del_a$}
\place{4.2}{1.45}{\large$x$}
\place{3.05}{2.0}{\large$\d x$}
\place{4.2}{2.53}{\large$\chi_{am}{}^n\,e_n$}
\place{0.7}{4.8}{\large$-\chi^a{}_{mn}\,e_a$}
\vskip-10pt
\capt{6.5in}{fig:prextrinsic}{The calculation of the extrinsic curvature for the fibres $X_y$ for the case that $c_a{}^m{\;=\;}0$.}
}
\end{figure}
\newpage
\setlength{\fboxsep}{0.5in}
\setlength{\fboxrule}{1pt}
\begin{figure}[H]
\centering
\vbox{
\framebox{\parbox{5.45in}{\vskip-5pt\hskip1.5cm\includegraphics[width=3.5in]{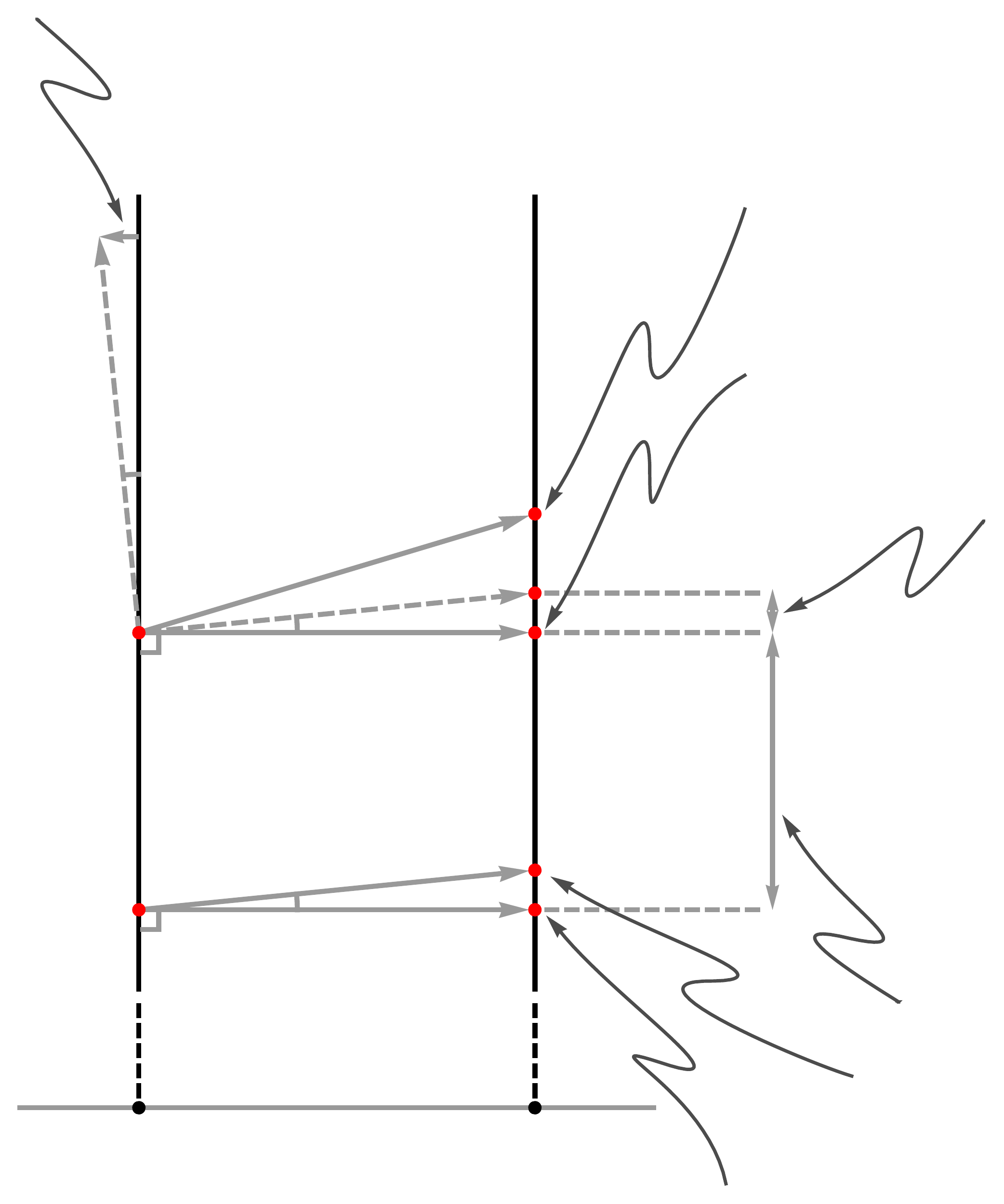}\hskip0.9in\vskip-10pt}}
\vskip29pt
\place{0.5}{5.05}{\large$ - \chi_{\a\,\mb\nb}\,=\, c_{\a\,(\mb;\nb)}$}
\place{1.35}{1.76}{\large$x$}
\place{0.95}{2.8}{\large$x+\d x$}
\place{1.55}{4.45}{\large$X_y$}
\place{1.55}{0.9}{\large$y$}
\place{2.9}{4.45}{\large$X_{y+\d y}$}
\place{2.77}{0.94}{\large$y+\d y$}
\place{0.8}{1.1}{\large$M$}
\place{3.7}{4.4}{\large$x+\d x$}
\place{3.8}{3.7}{\large$x+\d x - (c_\a+ \d c_\a) \d y^\a$}
\place{4.17}{1.15}{\large$x$}
\place{3.63}{0.75}{\large$x - c_\a\d y^\a$}
\place{2.15}{2.05}{\large$\d y^\a \del_\a$}
\place{2.1}{1.6}{\large$\d y^\a e_\a$}
\place{4.35}{1.45}{\large$\d x - \d c_\a \d y^\a$}
\place{4.6}{3.15}{\large$\chi_{\a\,\mb\nb}\,=- c_{\a\,(\mb;\nb)}$}
\vskip-10pt
\capt{6in}{fig:xtrinsic}{The calculation of the extrinsic curvature for the fibres $X_y$ for the case of complex  coordinates.}
}
\end{figure}

The extrinsic curvature is a tensor, as is $\D_{\a\,(\mb\nb)}$, so we see from \eqref{ChiInTermsOfC} that we cannot set $c_\a{}^\n{\=}0$. In fact the extrinsic curvature is the obstruction to so doing. The reason that this is so, is that the shift is defined in terms of the normal to the fibres, and so by the metric $\Ig_{mn}$, thus $c_\a{}^\n$ is not a holomorphic function of the coordinates and so cannot be removed by a holomorphic coordinate transformation.
\newpage                                      
\section{The universal  geometry $\ccU$}
\vskip-10pt
In the previous sections we have described in detail how to extend the geometry of $X$ to the larger structure of the fibration $\IX$. This also allowed us to describe geometrically the variations of the metric and complex structures on $X$ in terms of Lie derivatives and flows on the moduli space $M$. We now study the geometry $\ccU$, the universal bundle, whose base manifold is $\IX$. This is a holomorphic bundle with connection $\IA$, with $\IA$ the natural extension of $A$.  The field strength $\IF$ for $\IA$ has a tangibility $[1,1]$ part which exactly describes the variation of $A$. The Bianchi identity for $\IF$ efficiently encapsulates otherwise subtle identities derived in \citeI. 

The universal geometry also includes the three-form $H = \dd^c \o$ and its Bianchi identity \eqref{eq:Anomaly0}.
The extension of $H$ to $\IX$ is defined in a natural way 
\beq
\IH = \Id^c \Iomega~.
\label{eq:extendedsusy}\eeq
We demand that $\IH$ obeys an extended Bianchi identity
\beq
\Id \IH ~=- \frac{\ap}{4} \Big( \tr (\IF^2) - \tr (\IR^2)\, \Big)~.
\notag\eeq
Remarkably, this equation elegantly captures otherwise complicated algebraic relations derived with much effort in \citeI. These identities are important as they are central to the construction of the metric on $M$ and showing that it is \K. Using the extended quantities on $\IX$ we re-derive the metric on $M$ in a concise fashion in section \sref{s:metric}.
\subsection{The extension of $A$}
\vskip-10pt
The covariant derivative for $A$ defined in \citeI transforms covariantly under gauge transformations. It needs to be generalised to transform, additionally, under bundle diffeomorphisms \eqref{eq:BundleDiff}. To do this we define an extended connection $\IA$ for the extended vector bundle $\ccU \to \IX$
\beq\notag
\IA ~=~ A_m e^m + A^\sharp_a \dd y^a~, \qquad A^\sharp_a ~=~ \L_a - A_m c_a{}^m~,
\eeq
where the components of the corpus $A_m$ are identified with the connection along $X$.   In the following, we will denote the corpus of $\IA$ by $A = A_m e^m$ in the e-basis, the animus by $A^\sharp = A^\sharp_a \dd y^a$.  We can  divide the form into holomorphic type
\beq\notag
 \IA~=~\IcA - \IcA^{\dagger}~,\quad \IcA~=~ \IA^{(0,1)}~.
\eeq
We will not be specific about the structure group of the universal bundle $\ccU$ beyond requiring it contain $\eu{G}$ as a subgroup when restricted to $X$ appropriately. This restriction is important in later sections when we discuss deformations of $\ccT_X$. 

The form $\IcA$ can be decomposed into its animus and corpus\beq\notag
\begin{split}
 \IcA~=~ \cA^{\sharp}_{\ab}\,\dd y^{\ab}+\cA_{\mb}\,e^{\mb}~.
\end{split}\notag\eeq
The field strength of $\IA$ is defined as usual 
\beq
 \IF~=~ \Id\IA+\IA^2~.
 \label{eq:DoubleBarF}\eeq
This can be decomposed according to tangibility  and in terms of the covariant derivatives 
$\Deth, \Deth^\sharp$,  defined in  \eqref{eq:ExteriorDef} and \eqref{eq:dSharp}, respectively:
\beq\label{eq:IF}
 \IF~=~ (\Deth + \Deth^\sharp - S)(A+{A}^\sharp) + (A+{A}^\sharp)^2  ~=~ \half F_{mn} e^m e^n  +\dd y^a\,\IF^{\sharp}_{a} + \frac{1}{2}\,\IF^\sharp_{ab}\,\dd y^a \dd y^b~.
\eeq
Let us unpackage each of the three components of $\IF$. The corpus is  the field strength of $A$ on $X$, 
\beq
F_{mn}~=~ \del_m  A_n - \del_n A_m +A_m A_n - A_n A_m~.
\notag\eeq 
The second term defines a covariant derivative that transforms homogeneously under gauge transformations and is invariant under bundle diffeomorphisms: 
\beq
 \IF^{\sharp}_{a}\= \fD_a A~, ~~~\text{where} ~~~ \fD_a A~=~ e_a(A)-(\Deth c_a{}^m)A_m-\Deth_A A^\sharp_a~,\\[5pt] 
 \label{eq:MixedPart}\eeq
here 
\beq
\Deth_A A^\sharp_a \= \Deth A^\sharp_a + [A,A^\sharp_a]~,
\notag\eeq
and
\beq
e_a(A) \= \del_a A - c_a{}^m \del_m A~.
\notag\eeq 
On a gauge neutral object, $\fD_a$ reduces to $\Dethsharp_a$.

In holomorphic coordinates, using the identification of $\D_\a$ in \eqref{Deltaisdelc}, we find it is the appropriate generalisation of the holotypical derivative introduced in \citeI: 
\beq\notag
 \fD_{\a}\A \= e_{\a}(\A)-\D_{\a}{}^\n\,\A^\dag_\n-\dethb_\A A^\sharp_\a~.
\eeq
The third equation of \eqref{eq:IF} is
\beq
\IF^{\sharp}_{ab}\=2\,\Deth^\sharp_{[a} A^\sharp_{b]}+[A^\sharp_a,A^\sharp_b]-S{}_{ab}{}^m A_m~,~~~\text{where}~~~\Dethsharp_a A^\sharp_b=e_a(A^\sharp_b)~.
\notag\eeq

We take $\ccU$ to be holomorphic meaning  
\beq
\IF^{(0,2)}\= 0~.
\notag\eeq 
The corpus of $\IF$ automatically satisfies this requirement in virtue of $F^{(0,2)} {\=} 0$. The tangibility $[1,1]$ component is the condition that $\A$ depend holomorphically on parameters
\beq\notag
\fD_{\bb}\cA\= 0~.
\eeq
The tangibility $[2,0]$ component implies $\IF^\sharp_{\ab\bb}{\=}0$. That is, that the bundle $\ccU$ restricted to $M$ is holomorphic. In deducing this we have used $S_{\ab\bb} = 0$.

 Consider now the Bianchi identity for $\IF$ 
\beq
\Id_\IA \IF ~=~ 0~.
\label{bianchi1}\eeq
The corpus realises the Bianchi identity on $X$. The animus gives two further identities
\beq\label{BidIF2}
 \begin{split}
 \Deth_A(\fD_a A)\= \fD_a F~~~\text{and}~~~
  [\fD_a,\fD_b]A\;=\, -\Deth_A(\IF^{\sharp}_{ab})+S{}_{ab}{}^m\,F_{m}\= 0~,
\end{split}\notag\eeq
where
\begin{align*}
 \Deth_A(\fD_a A) &\= \Deth(\fD_a A)+[A,\fD_a A]~,&
 \fD_a F &\=  \Dethsharp_a F+[A^\sharp_a,F]~,\\[0.2cm]
 \fD_a (\fD_b A) &\=  \Dethsharp_a(\fD_{b}A)+[A^\sharp_a,\fD_b A]~,&
 \Deth_A \IF^{\sharp}_{ab} &\= \Deth\IF^{\sharp}_{ab}+[A,\IF^{\sharp}_{ab}] ~.
\end{align*}
The relations \eqref{BidIF2} can be derived directly from the definition of the covariant derivative as in~\citeI with some labour. What we see here is an alternative derivation through the Bianchi identity. This also has the advantage of unification, reducing a pair of identities to a single identity.  

The Atiyah constraint comes from taking $a{\=}\a$ in the first equation of \eqref{BidIF2},  and considering the $(0,2)$-component together with the identification of $\D_\a$ in \eqref{Deltaisdelc}:
\beq\notag
 \dethb_{\cA}(\fD_\a\cA)~=~ \D_\a{}^\m\, F_{\m}~.
\eeq
\subsection{The extension of $B$ and $H$}
\vskip-10pt
The field $\IH$ is the extension of $H$, and defined as
\beq
\IH ~=~ \Id \IB - \frac{\ap}{4} \Big(\CS[\IA] - \CS[\ITheta]\Big)~,~~~\text{where}~~~
\CS[\IA]~=~\tr\!\left(\IA\,\Id \IA +\smallfrac23 \IA^3\right)~,
\label{eq:IHdef3}\eeq
where $\IB$ is the extension of the Kalb--Ramond field
\beq\notag
\IB \= \frac12 B_{mn}\,e^m e^n +\IB^\sharp_{am}\, \dd y^a e^m  +
\frac12 \IB^\sharp_{ab}\,\dd y^a\dd y^b \= B + \IB^\sharp_a\, \dd y^a + {B}^\sharp~ .
\eeq
$\IH$ decomposes as
\beq
 \IH~=~\frac{1}{3!}\,\dd y^{abc}\,\IH\#_{abc}+\frac{1}{2}\,\dd y^{ab}\,\IH\#_{ab}+\dd y^a\,\IH_a\#+H\ ,
\notag\eeq
where the $[1,2]$ term will be relevant in what follows.  It is given by
\beq\label{tang12IH}
 \IH\#_{a} = \Dethsharp_a B - \Deth\IB\#_{a} - 
 \frac{\ap}{4}\Big(\tr(A\#_a\,\Deth A) - \tr{(\Th\#_a\,\Deth \Th)}\Big) +
\frac{\ap}{4}\Big(\tr(A\,\fD_a A) - \tr{(\Th\,\fD_a \Th)}\Big)\; .
\eeq
We can now rewrite this in terms of  covariant derivatives 
\beq\label{curlyB}
\IH^\sharp_a \= \fD_a B+\frac{\ap}{4}\,\Big(\tr{(A\,\fD_a A)}-\tr{(\Th\,\fD_a \Th)}\Big)-\Deth\IB\#_a ~,
\eeq
with the covariant derivative $\fD_a B$ is defined as 
\beq
\fD_a B\=\Dethsharp_a B-\frac{\ap}{4}\Big(\,\tr(A\#_a\,\Deth A)-\tr{(\Th\#_a\,\Deth \Th)}\Big)\ ,
\notag\eeq
which sharpens the relation derived in \citeI. We will see why this is a covariant 
derivative shortly.

By demanding $\IH$ be gauge invariant, we see that the field $\IB$ transforms under gauge transformations:
\beq
\IB \to {}^{\Ph,\Ps} \IB\= \IB + \frac{\ap}{4}\Big\{\tr\big( \IY \IA - \IZ\, \ITheta  \big) + \IU - \IW \Big\} ~.
\label{eq:GaugedB}\eeq
Which is the natural extension of the rule given in \eqref{eq:BTransf2}.
Given the above relations the field strength $\IH$ is invariant. As the animus of $\IB$ transforms inhomogeneously, it is inconsistent to try to set it to zero. 
Here $\IY$, $\IU$ are the extensions of $Y$ and $U$:
\beq\begin{split}
 \IY~=~\Ph^{-1}\,\Id\Ph\  &,\ \ \ \Id\IU\= \frac{1}{3}\,\tr\IY^3\ ,
\end{split}\notag\eeq
with $\IZ,\IW$ being the spin connection counterpart.

The right hand side of \eqref{curlyB} is the combination of terms identified in \citeI as being gauge invariant. This we now understand since $\ccB_a {\=} \IH_a^\sharp$ and $\IH$ is gauge invariant.

The covariant derivative is defined such that it transforms in a manner parallel to the $B$-field itself:
\beq
 {}^{(\Ph,\Ps)}\fD_a B~=~\fD_a B+\frac{\ap}{4}\Big(\tr{(Y\,\fD_a A)}+\mathfrak{Y}_a-\tr{(Z\,\fD_a\Th)}-\mathfrak{Z}_a\Big) .
\notag\eeq
We have also defined
\beq\begin{split}
 \mathfrak{Y}_a& \= \Dethsharp_a U-\tr{(Y\#_a Y^2)}+\Deth\left(\tr{(Y\#_a A-A Y\#_a)}\right)~,\\[0.2cm]
 \mathfrak{Z}_a& \= \Dethsharp_a Z-\tr{(Z\#_a Z^2)}+\Deth\left(\tr{(Z\#_a \Th-\Th\#_a Z)}\right)\ ,
\end{split}\notag\eeq
Using that the form $\IY$ satisfies $\Id\IY=-\IY^2$, we find that this quantity is $\Deth$-closed
\beq
 \Deth\mathfrak{Y}_a\= 0~.
\notag\eeq

In addition to the gauge transformations above the field strength $\IH$ is invariant under an additional  symmetry, in which $\IB$ shifts by a  $\Id$-exact amount,
\beq
 \IB\to\IB+\Id\Ibeta\ ,\ \ \ \Ibeta=\b_m\,e^m+\b^\sharp_a\,\dd y^a\ ,
\notag\eeq
where the one-form $\Ibeta$ is gauge-invariant. Decomposing this into tangibilities we have
\beq\begin{split}
 &B\to B+\Deth\b\ ,\\[0.1cm]
 &\IB^\sharp_a\to\IB^\sharp_a+\Dethsharp_a\b-\Deth\beta^\sharp_a\ ,\\[0.1cm]
 &\IB^\sharp_{ab}\to\IB^\sharp_{ab}+\Dethsharp_a\beta^\sharp_b - 
 \Dethsharp_b\beta^\sharp_a-S{}_{ab}{}^m\,\b_m\ .
\end{split}\label{eq:IBshift}\eeq
The first line corresponds to shifting $B$ by a $\Deth$-exact term. The second line corresponds to shifts of $\IB_a^\sharp$. The way to think of $\IB^\sharp_a$ is that it is another connection; its purpose to is define an invariant quantity $\ccB_a$ as in  \eqref{curlyB}. This invariance can be checked directly, but an easier way to see this is to note that $\IH$ is invariant and so $\IH_a^\sharp = \ccB_a$ is invariant.   The quantity~$\ccB_a$, mentioned in the introduction, plays an important role as $\ccB_a  + \ii\fD_a \o$ plays the role in heterotic geometry analogous to the role of complexified \K class in special geometry.  All this goes to show that the animi of $\IA$ and $\IB$ are connections which are needed to define covariant derivatives on the moduli space.

Although we have not fully explored this aspect, we believe the quantity  $\IB^\sharp_{ab}$ with the transformation rules as in the third line above, provide connections that enable one to define second and higher order derivatives. For example, see \citeI where a second order covariant derivative was defined.

\subsection{The extension of $\dd^c \omega$}\label{subsec:dcomega}
\vskip-10pt
We will shortly have need for the quantity 
$$
\Id^c\Iomega \=\frac{1}{3!} \IJ^P \IJ^Q \IJ^R (\Id \Iomega)_{PQR}~.
$$
In a holomorphic basis $\Iomega$ is $(1,1)$ and so
$$
\Id^c\Iomega \= \ii (\Id \Iomega)^{(2,1)} - \ii (\Id\Iomega)^{(1,2)}~.
$$
The term $\Id^c\Iomega$ has vanishing $[3,0]$ term due to the fact that $g\#_{\a\bb}$ is K\"ahler, while the remaining components are given by
\beq\label{Idcom}
\begin{split}
(\Id^c\Iomega)_{\a\phantom{\b}} \=&~ \ii\,\fD_\a\o^{(1,1)}-\ii \fD_\a \o^{(0,2)} ~,\\[0.3cm]
(\Id^c\Iomega)_{\a\b} \=& -\ii\,S{}_{\a\b}{}^\m\,\o_\m~,
\hskip40pt (\Id^c\Iomega)_{\ab\bb}\= \ii\,S{}_{\ab\bb}{}^\mb\,\o_\mb~ ,\\[0.3cm]
(\Id^c\Iomega)_{\a\overline\b} \=& -\ii\,S{}_{\a\bb}{}^{\mb}\,\o_{\mb}+
\ii\,S{}_{\a\overline\b}{}^\m\,\o_\m\ .
\end{split}\eeq
Note that the action of the covariant derivative $\fD_\a$ on a gauge neutral object is the same as $\Dethsharp$ so that $\fD_\a \o^{(p,q)} = \Dethsharp_a\o^{(p,q)}$. In the sections to follow, where no ambiguity will arise we will use $\fD_\a$ to prevent an unnecessary proliferation of symbols. 

On setting $S=0$ the expression simplifies significantly 
\beq
\Id^c\Iomega \= \ii (\deth - \dethb) \omega + \ii \dd y^\a (\fD_\a \omega^{(1,1)} - \fD_\a \omega^{(0,2)}) + \ii \dd y^\bb (\fD_\bb \omega^{(2,0)} - \fD_\bb \omega^{(1,1)}) ~.
\notag\eeq
While $\omega$ is type $(1,1)$, its derivative  $\fD_\a$ is type $(2,1)\oplus (1,2)$: $\fD_\a \omega {\=} \fD_\a\omega^{(1,1)} + \fD_\a\omega^{(0,2)}$, and this expresses the type changing property of variations with respect to complex structure.

\subsection{The relation $\IH = \Id^c \Iomega$, Bianchi identity and second order relations}
\vskip-10pt
We suppose that the extended supersymmetry relation \eqref{eq:extendedsusy} holds on $\IX$
This imposes some constraints on the variations of a heterotic structure. The tangibility $[1,2]$ part of this relation gives
\beq\begin{split}\label{susyoncF0}
 \ccB_\a{}^{(2,0)}~&=~  0 ~,\\[0.2cm]
 \ccB_\a{}^{(1,1)}-\ii\,\fD_{\a}\o^{(1,1)}~&=~ 0~,\\[0.2cm]
 \ccB_\a{}^{(0,2)}+\ii\,\fD_\a \o^{(0,2)}  ~&=~ 0~.
\end{split}\eeq
We define 
\beq\notag
\ccZ_\a \= \ccB_\a + \ii \fD_\a \o~, \quad \text{and} \quad \ccZb_\a \=   \ccB_\a - \ii \fD_\a \o~,
\eeq
which are the generalisation to heterotic geometry of the variation of the complexified \K class familiar in special geometry $\d B + \ii \d \o$. 
In terms of $\ccZ, \ccZb$, \eqref{susyoncF0} can be written as 
\beq\begin{split}\label{susyoncF}
& \ccZ_\a{}^{(2,0)}~=~ \ccZb_\a{}^{(2,0)} ~=~  0 ~,\\[0.2cm]
 &\ccZb_\a{}^{(1,1)} ~=~ 0~,\\[0.2cm]
 &\ccZ_\a{}^{(0,2)}  ~=~ 0~.
\end{split}\eeq
These equations described first order conditions on the heterotic moduli which were derived in \cite{Anderson:2014xha,delaOssa:2014cia,delaOssa:2014msa,Garcia-Fernandez:2018ypt} and in this notation in \citeI by taking partial derivatives of the supersymmetry relation $H=\dd^c\omega$. We identify $\IB_a$ with $b_a$ and note that $\IH = \Id^c \Iomega$ captures all of the moduli equations except one.  For the remaining one we turn to the Bianchi identity for $\Id\IH$ on $\IX$: 
\beq
\Id\IH~=-\frac{\ap}{4}\Big(\tr{\IF^2}-\tr{\IR^2}\Big) \=  \Id( \Id^c\Iomega)\ .
\notag\eeq
The curvatures $\IF$ and $\IR$ are of type $(1,1)$ and so only the type $(2,2)$ part of this relation is non-vanishing.

We start with tangibility $[1,3]$, focusing on holomorphic variation with index $\a$. The first equality of the previous equation is 
\beq\label{Bid13}
 (\Id\IH)_\a~=-\frac{\ap}{2}\Big(\tr{(\fD_\a\cA\ F)}-\tr{(\fD_\a\vth\ R)}\Big)\ .
\notag\eeq
Meanwhile $(\Id \Id^c \Iomega)_\a$  is simplified using
\beq
\begin{split}
 \Deth (\Id^c \Iomega)_\a ~&=~ \ii \Deth (\fD_\a \o^{1,1} - \fD_\a \o^{0,2})~, \\
   \fD_\a(\Deth^c\o)~&=~2\ii\,\D_\a{}^\m\,(\deth\o)_\m-2\ii\,\deth(\D_\a{}^\m\,\o_\m) +
    \ii (\deth - \dethb) \fD_\a \o\ ,
\end{split}\notag\eeq
and by using \eqref{susyoncF0} we get
\beq\begin{split}
 \dethb( \ccZ_\a^{(1,1)})&\= 2\ii\,\D_\a{}^\m\,(\deth\o)_\m+\frac{\ap}{2}\Big(\tr{(\fD_\a\cA\ F)}-\tr{(\fD_\a\vth\ R)}\Big)~. 
\end{split}\label{susyoncF2}\eeq

Let us now turn our attention to tangibility $[2,2]$. Assuming that $S = 0$, this consists of two relations
\beq\label{Bianchitang22}
\begin{split}
\fD_\a(\Id^c\Iomega)_\b-\fD_\b(\Id^c\Iomega)_\a~&=~-\frac{\ap}{2}\,\Big(\tr{(\fD_\a\cA\,\fD_\b\cA)}-\tr{(\fD_\a\vth\,\fD_\b\th)}\Big)\ ,\\[0.6cm]
\fD_\a(\Id^c\Iomega)_{\bb}-\fD_{\bb}(\Id^c\Iomega)_\a~&=~-\frac{\ap}{2}\,\Big(\tr{(\fD_\a\cA\,\fD_{\bb}\cA^\dagger)}-\tr{(\fD_\a\th\,\fD_{\bb}\th^\dagger)}\Big)\\
&\hskip3cm -\frac{\ap}{4}\Big(\tr{(\IF\#_{\a\bb}\,F)}-\tr{(\IR\#_{\a\bb}\,R)}\Big)\ .
\end{split}
 \raisetag{0cm}
\eeq
The second relation forms part of a critical algebraic relation needed to derive the moduli space metric in \citeI and so we focus on this one.
It becomes 
\beq\begin{split}
 \fD_\a(\Id^c\Iomega)_{\bb}-\fD_{\bb}(\Id^c\Iomega)_\a &\=-2\ii\,\big(\fD_\a \fD_{\bb}\o\big)^{(1,1)}+2\ii\,\D_\a{}^\m\,(\fD_{\bb}\o^{(2,0)})_\m+2\ii\,\D_{\bb}{}^{\nb}\,(\fD_\a\o^{(0,2)})_{\nb}\\[0.2cm]
 &\hskip-10pt +\ii\,\fD_\a(\D_{\bb}{}^{\nb}\,\o_{\nb})-\ii\,\D_{\bb}{}^{\nb}(\fD_\a\o^{(1,1)})_{\nb}+\ii\,\fD_{\bb}(\D_\a{}^\m\,\o_\m)-\ii\,\D_\a{}^\m(\fD_{\bb}\o^{(1,1)})_\m\ .
\end{split}\notag\eeq
\vskip-20pt
The last equation can be simplified by noticing a further relation 
\beq
 \fD_\a(\D_{\bb}{}^{\nb}\,\o_{\nb})-\,\D_{\bb}{}^{\nb}(\fD_\a\o^{(1,1)})_{\nb}\=(\deth S{}_{\a\bb}{}^{\nb})\,\o_{\nb}\=0\ ,
\notag\eeq
which sets the last line to zero. Putting everything together, we can rearrange (\ref{Bianchitang22}) to~obtain
\beq\begin{split}\label{secondorderom}
  &\big(\fD_\a\fD_{\bb}\o\big)^{(1,1)}~=~-\frac{\ii\ap}{4}\,\Big(\tr{(\fD_\a\cA\,\fD_{\bb}\cA^\dagger)}-\tr{(\fD_\a\th\,\fD_{\bb}\th^\dagger)}\Big)\\[0.2cm]
  &\hspace{1cm}-\frac{\ii\ap}{8}\Big(\tr{(\IF\#_{\a\bb}\,F)}-\tr{(\IR\#_{\a\bb}\,R)}\Big)+\D_\a{}^\m\,(\fD_{\bb}\o^{(2,0)})_\m+\D_{\bb}{}^{\nb}\,(\fD_\a\o^{(0,2)})_{\nb}\ .
\end{split}\notag\eeq
This shows that the Bianchi identity for $\IH$ incorporates the second order algebraic relation for the variation of the hermitian form that is crucial in deriving the $\ap$-corrected moduli~metric.  
\newpage
\section{Using $\IX$ to deform connections on $\ccT_X$}
\vskip-10pt
The moduli space metric, expressed in \eqref{eq:modulimetric}, has a contribution from the variation of $\Th$.
As is the case for the gauge connection $A$, the variation of $\Th$ is expressed as a covariant derivative with respect to parameters. However, unlike $A$, the derivatives of $\Th$ are tied to the geometry of $X$, up to Lorentz gauge transformations. Our aim in this section is to compute $\fD_\a \Th$ to zeroth order in $\ap$, expressing the answer in terms of the moduli of $X$. 

\subsection{The covariant derivative of  $\Th$}
\vskip-10pt
Our approach to computing  covariant derivatives of $\Th$ is to extend this connection to $\IX$. The connection $\ITheta$ is then a connection on the frame bundle, and its associated tangent bundle  $\ccT_\IX$. We take the connection $\ITheta$ to be metric compatible. 
It has  a curvature two-form $\IR  = \Id \ITheta + \ITheta^2$ which transforms in a Lorentz algebra $\so(D)$ where $D = \dim \IX$. In terms of tangibilities,  $\ITheta$ decomposes as
\beq\label{decomposeITh}
 \ITheta\= \Th^\sharp_a\,\dd y^a+\Th_m\,e^m~. 
\eeq
The curvature $\IR$  has a  tangibility $[1,1]$ component which defines the covariant derivative 
\beq\notag
 \fD_{a}\Th\= \IR^\sharp_{am} e^m ~.
\eeq

For any  frame on $\IX$ with basis of sections $\{s_a,s_m\}$, the connection $\Th$ has  symbols
\beq\notag
 \Inabla s_a\= \ITheta^b{}_a\,s_b+\ITheta^n{}_a\,s_n~,\hskip30pt 
 \Inabla s_m\= \ITheta^n{}_m\,s_n+\ITheta^b{}_m\,s_b~.
\eeq

In the physical string theory, the connection $\Th$ is $\so(6)$ valued, when $(s^a,\,s^m)$ form an orthonormal basis, and the term that appears in the moduli space metric involves a trace over~$\so(6)$: 
\beq\notag
\tr\Big(\fD_\a\Th\star\fD_{\bb}\Th\Big)\= (\fD_\a\Th)^m{}_n \star(\fD_{\bb}\Th)^n{}_m~.
\eeq
However, under a change of basis 
\beq\label{changebasis}
 s_a=s^b{}_a\,e_b~,\quad s_m=s^n{}_m\,\del_n~,
\eeq
for some invertible matrices $s^b{}_a$ and $s^n{}_m$. The covariant derivative $\fD_a \Th$ transforms as
\beq
\fD_\a  \Th^m{}_n\to s^m{}_k \,\fD_\a \Theta^k{}_l\,s^{-1}{}^l{}_n~,
\notag\eeq
and so $\tr (\fD_\a\Th\star\fD_{\bb}\Th)$ is invariant.  Hence, we are free to compute this term in the $e$-basis, which turns out to be very convenient. Our first task then is to compute the covariant derivative, 
\beq\label{varGamma}
 \fD_{a}\Th^n{}_m \= 
 \Dethsharp_a\Th^n{}_m-\Deth\Th^\sharp_a{}^n{}_m +[\Th^\sharp_a,\Th]^n{}_m~,
\eeq
in the $e$-basis. The first term in \eqref{varGamma} is evaluated like a 1-form using \eqref{eq:delsharpeta}. In explicit detail it is given by $\Dethsharp_a \Th^n{}_m{\=}e_a (\Th^n{}_m) - e^p (\del_p c_a{}^q) \Th_q{}^n{}_m $, while the second term is 
$\Deth \Th_a^{\sharp\, n}{}_m {\=} e^m \del_m \Th_a^{\sharp\, n}{}_m$. 

\subsection{A two-parameter family of connections \smash{$\ITheta^{(\e,\r)}$ on $\IX$}}
\vskip-10pt
As reviewed in Appendix \ref{s:HetGeometry},   the supersymmetry Killing spinor of heterotic supergravity is covariantly constant with respect to the connection
$$
\Th_m^\Bi \= \Th_m^{\LC} - \half H_m~.
$$ 
By using this and writing $J$ as a spinor bilinear, it follows that  $J$ is covariantly constant with respect to this connection $\nabla^\Bi J = 0$, and so $\Th^\Bi$ is hermitian. Furthermore, the equation $\nabla^\Bi J = 0$ can be expressed in terms of forms, and when so written,  in a holomorphic frame, we have that 
$H {\=} \ii (\del - \delb)\o$, as we see from \eqref{eq:Jcovconstant} and \eqref{eq:Jcovconstant2}. This is stated more generally as $H {\=} \dd^c \o$. On the other hand the torsion of $\Th^\Bi$, as defined in \eqref{eq:torsion}, is exactly $H$,  more precisely  $T^m{}_{np} {\=} H^m{}_{np}$. Thus, we find that
$\Th^\Bi$ has totally antisymmetric torsion equal to~$\dd^c \o$. This connection is known in the mathematics literature as the Bismut connection. 

While the supersymmetry spinor is covariantly constant with respect to $\Th^\Bi$,  a different connection 
$\Th^\Hu$ appears in the heterotic action \eqref{eq:10daction}. It  is non-hermitian and has torsion given by $-H$. Hence, 
$$
\Th_m^\Hu \= \Th_m^{\LC} + \half H_m~.
$$
We call this the Hull connection. 

The last term in the moduli space metric \eqref{eq:modulimetric} derives from dimensionally reducing the quantity $\tr |R(\Th^\Hu)|^2$, and our task therefore is to compute the covariant derivative of the Hull connection $\fD_a \Th^\Hu$. However, we will first work in  more generality and compute the covariant derivative of a two-parameter family of connections introduced in \cite{2017Otal}, of which $\Th^\Bi, \Th^\Hu$ are special cases. This family also includes the 1-parameter family defined by Gauduchon \cite{1997Gauduchon}. 
We will show that only a 1-parameter subfamily are holomorphic on~$\IX$, and this includes the Hull connection.  To define the family, consider a fixed complex manifold $X$, and on $\ccT_X$ we introduce the connection $\Th^{(\e,\r)}$, with $\e,\r \in \IR$ and  symbols
\beq\label{eq:GammaFamily}
 \begin{split}
  \Th^{(\e,\r)}{}_\m{}^\n{}_\s 
  &\=   \Th^{\LC}{}_\m{}^\n{}_\s+\frac{(\e-\r)}{2}\,H_{\m}{}^\n{}_\s~,\\[0.1cm]
  \Th^{(\e,\r)}{}_\mb{}^\n{}_\s 
  &\=     \Th^{\LC}{}_\mb{}^\n{}_\s+\frac{(\e-\r)}{2}\,H_{\mb}{}^\n{}_\s~,\\[0.1cm]
  \Th^{(\e,\r)}{}_\m{}^\nb{}_\s 
  &\=  0~,\\[0.1cm]
  \Th^{(\e,\r)}{}_\mb{}^\nb{}_\s 
  &\=     \Th^{\LC}{}_\mb{}^\nb{}_\s+\frac{(\e+\r)}{2}\,H_\mb{}^\nb{}_\s\,~,
 \end{split}
\eeq
where $\Th^{\LC}$ is the Levi--Civita connection and $H {\=} \dd^c\o$. The Bismut connection is given by
$\Th^\Bi {\=} \Th^{(-1,0)}$, the Hull connection by $\Th^\Hu {\=} \Th^{(1,0)}$ and the Chern connection by $\Th^{\Ch} {\=} \Th^{(0,-1)}$. Furthermore, when  $\r+\e{\;=}-1$ this reduces to the 1-parameter family of Gauduchon. 

To compute $\fD_a\Th^{(\e,\r)}$ we extend $\Th$ to $\IX$ as follows. Firstly, we consider the Levi--Civita connection $\ITheta^{\LC}$ on $\IX$. In terms of an arbitrary basis of vectors, denoted $e_P$, using \eqref{eq:ThLC} it  has symbols given by
\beq\label{eq:LCinframe}
\begin{split}
 &\ITheta^{\LC}_P{\,}^Q{}_R~=~ \frac{1}{2}\,\Ig^{QS}\Big(e_P(\Ig_{SR})+e_R(\Ig_{SP})-e_S(\Ig_{PR})\Big)\\
 &\hspace{2.5cm}-\frac{1}{2}\,\Ig^{QS}\Big([e_P,e_S]^T\,\Ig_{TR}+[e_R,e_S]^T\,\Ig_{TP}\Big)+\frac{1}{2}\,[e_P,e_R]^Q~.
\end{split}
\eeq
For the e-basis $e_P = (e_a,\del_m)$ we get
\beq\label{eq:LConIX}
 \begin{split}
\ITheta^{\LC}{\,}^{n}{}_{k}
&\= \+ e^m\,\G_m^{\LC}{}^n{}_k+\dd y^a\Big(\del_k c_a{}^n+
  \frac{1}{2}\,g^{nl}\fD_a\,g_{lk}\Big)~,\\[0.2cm]
\ITheta^{\LC}{\,}^b{}_{k} 
&\= - \frac{1}{2}\,e^m\,g^\sharp{}^{bd}\fD_d\,g_{mk}-
  \frac{1}{2}\,\dd y^a\,g^\sharp{}^{bd}\,S{}_{ad}{}^l\,g_{lk}~,\\[0.2cm]
\ITheta^{\LC}{\,}^n{}_{c} 
&\= \+\frac{1}{2}\,e^m\,g^{nl}\fD_{c}\,g_{lm}+\frac{1}{2}\,\dd y^a\,S{}_{ac}{}^n~,\\[0.2cm]
\ITheta^{\LC}{\,}^b{}_{c} &
\= -\frac{1}{2}\,e^m\,g^\sharp{}^{bd}\,S{}_{cd}{}^l\,g_{lm}+
  \dd y^a\,\G^\sharp{}^{\LC}{}_{ac}{}^b~,
\end{split}\eeq
where $\G^{\LC}$ and $\G^{\sharp\LC}$ are the standard expressions, see \eqref{eq:GLC}, in terms of $g_{mn}$ and $g^\sharp_{ab}$ respectively;  
$\fD_a g_{mn} {\=} (\Dethsharp_a g)_{mn} {\=} (\cL_{e_a} g)_{mn}$ and  \eqref{varxicomponents} provides us with an expression in components
\beq\notag
\begin{split}
\fD_a\,g_{mn}
&\= e_a(g_{mn}) - c_a{}^k{}_{,\,m}\, g_{kn} - c_a{}^k{}_{,\,n}\, g_{mk} ~.
\end{split}\raisetag{18pt}
\eeq
The connection $\ITheta^{(\e,\r)}$  extends \eqref{eq:GammaFamily} to $\IX$ in a natural way. The resulting symbols  are written below in holomorphic coordinates.  We have used \eqref{eq:LConIX},  $\IH = \Id^c \Iomega$ and the calculation of $\Id^c \Iomega$ in \eqref{Idcom}.

\begin{align*}
\intertext{$\bullet$~~Internal indices purely vertical: }
  \ITheta^{(\e,\r)}{\,}^\n{}_\s 
  &\=e^{\m}\ \Th^{(\e,\r)}_\m{}^\n{}_\s+e^{\mb}\ \Th_\mb^{(\e,\r)}{}^\n{}_{\s}+
  \dd y^\a\ \Big(\del_\s c_\a{}^\n+\frac{(1+\e-\r)}{2}\ g^{\n\lb}\,\fD_\a g_{\s\lb}\Big)+\\
  &\hskip50pt +\frac{(1-\e+\r)}{2}\,\dd y^{\ab}\ g^{\n\lb}\,\fD_{\ab}g_{\s\lb}~,\\[10pt]
  \ITheta^{(\e,\r)}{\,}^\nb{}_\s
  &\=e^{\mb}\ \Th^{(\e,\r)}_\mb{}^\nb{}_\s+(1+\e+\r)\,\dd y^{\ab}\ g^{\nb\l}\,\D_{\ab[\l\s]}~.
  \\[5pt]
\intertext{$\bullet$~~Internal indices of mixed type, upper index horizontal:}
  \ITheta^{(\e,\r)}{\,}^\b{}_\s
  &\= -e^{\m}\ g^\sharp{}^{\b\db}\Big(\D_{\db(\m\s)}+(\e-\r)\D_{\db[\m\s]}\Big)-
  \frac{(1-\e+\r)}{2}\,e^{\mb}\ g^\sharp{}^{\b\db}\,\fD_{\db}g_{\s\mb}+\\
  &\hskip50pt +\frac{(1+\e-\r)}{2}\,\dd y^\a\ g^\sharp{}^{\b\db}\,S{}_{\db\a}{}^{\lb}\,g_{\s\lb}+\frac{(1-\e+\r)}{2}\,\dd y^{\ab}\ g^\sharp{}^{\b\db}\,S{}_{\db\ab}{}^{\lb}\,g_{\s\lb}~,\\[10pt]
\ITheta^{(\e,\r)}{\,}^\bb{}_\s
&\= -\frac{(1+\e+\r)}{2}\,e^{\mb}\ g^\sharp{}^{\bb\d}\,\fD_{\d}g_{\s\mb}+
\frac{(1+\e+\r)}{2}\,\dd y^{\ab}\ g^\sharp{}^{\bb\d}\,S{}_{\d\ab}{}^{\lb}\,g_{\s\lb}~.
\\[5pt]
\intertext{$\bullet$~~Internal indices of mixed type, upper index vertical:}
  \ITheta^{(\e,\r)}{\,}^\n{}_\g
  &\= \frac{(1-\e+\r)}{2}\,e^\m\ g^{\n\lb}\,\fD_\g g_{\m\lb}+e^{\mb}\,g^{\n\lb}\Big(\D_{\g(\mb\lb)}+(\e-\r)\D_{\g[\mb\lb]}\Big)\,\\
  &\hskip50pt +\frac{(1-\e+\r)}{2}\,\dd y^\a\,S{}_{\a\g}{}^{\n}+
  \frac{(1+\e-\r)}{2}\,\dd y^{\ab}\ S{}_{\ab\g}{}^{\n}~,\\[10pt]
  \ITheta^{(\e,\r)}{\,}^\nb{}_\g
  &\= \frac{(1+\e+\r)}{2}\,e^{\mb}\ g^{\nb\l}\,\fD_\g g_{\l\mb}+
  \frac{(1+\e+\r)}{2}\,\dd y^{\ab}\ S{}_{\ab\g}{}^{\nb}~.\\[5pt]
\intertext{$\bullet$~~Internal indices purely horizontal:}
  \ITheta^{(\e,\r)}{\,}^\b{}_\g 
  &\= \frac{(1-\e+\r)}{2}\,e^{\m}\ g^\sharp{}^{\b\db}\,S{}_{\db\g}{}^{\lb}\,g_{\m\lb}+
  \frac{(1-\e+\r)}{2}\,e^{\mb}\ g^\sharp{}^{\b\db}\,S{}_{\db\g}{}^{\l}\,g_{\l\mb}+
  \dd y^\a\ \Th^{\sharp}{}_\a{}^\b{}_\g~,\\[10pt]
  \ITheta^{(\e,\r)}{\,}^\bb{}_\g
  &\= \frac{(1+\e+\r)}{2}\,e^{\mb}\ g^\sharp{}^{\bb\d}\,S{}_{\d\g}{}^{\l}\,g_{\l\mb}~.\\
\end{align*}
Some comments are in order. First, the symbols $\ITheta^{(\e,\r)}$ coincide with the symbols  $\Th^{(\e,\r)}$ on the fibre $X$ when all three indices are vertical. This is not the case if the symbols are expressed in the coordinate basis $\{\del_a,\del_m\}$. Second, when all indices are horizontal, $\ITheta$ coincides with $\Th^\sharp$ the connection symbols formed from $g^\sharp$. As $g^\sharp$ is \K the connection is the unique hermitian torsionless connection on $M$, whose only nonvanishing components are
\beq\notag
 \Th^\sharp{}_\a{}^\b{}_\g \=g^\sharp{}^{\b\db}\del_\a g^\sharp_{\db\g}~.
\eeq
Finally, we denote the split $\Inabla$ according to tangibility as follows
\beq
\Inabla \= \nabla + \nabla^\sharp \= e^m \nabla_m + \dd y^a \nabla^\sharp_a~.
\notag\eeq

\subsection{The covariant derivative of \smash{$\Th^{(\e,\r)}$}}
\vskip-10pt
We now compute $\fD_\a \Th^{(\e,\r)}$ to zeroth order in $\ap$. We will find that only when $\e-\r{\=}1$ is the connection holomorphic, that is $\IR^{(0,2)} = 0$. 
For the remainder of the paper we set $S_{ab} = 0$, and work in harmonic gauge, the conventional choice in supergravity: $\nabla^m \d g_{mn} {\=} 0$ where $\nabla_m {\=} \del_m + \Th_m$ is computed with respect to the affine spin connection on $X$ that is discussed in Appendix \ref{s:HetGeometry}. This gauge fixing decomposes into
\beq
 \D_\a{}^\m\,\o_\m\=0~,\ \ \ \nabla_\m\,\D_\a{}^\m\=0~,\ \ \ \del_m \big(\o^{\m\nb}\fD_\a\o_{\m\nb}\big)\=0~,
\notag\eeq
provided $X$ has $h^{(0,2)} = 0$. Interestingly, without vanishing curvature $S=0$ and gauge fixing, the connection is not holomorphic for any choice of $\e,\r$.

First, we demand that the connection is holomorphic $\ccD_\a \Th_\m = 0$. Using \eqref{varGamma}, we find the following components are not immediately zero:
\beq\notag
  \fD_{\a}\Th^{(\e,\r)}{\,}_\m{}^\n{}_\s \=
  \frac{(1-\e+\r)}{2\ii}\,g^{\n\lb}\ \nabla_\m\,\fD_\a \o_{\s\lb}~,
\quad
  \fD_{\a}\Th^{(\e,\r)}{\,}_\m{}^\nb{}_\sb\=
  \!-\frac{(1-\e+\r)}{2\ii}\,g^{\nb\l}\ \nabla_\m\,\fD_\a\o_{\l\sb}~. 
\eeq
We see that the covariant derivatives of the variations appear 
\beq\notag
  \nabla_\s\,\D_{\a\mb}{}^\n\=\del_\s\,\D_{\a\mb}{}^\n+\Th_\s{}^\n{}_\l\,\D_{\a\mb}{}^\l~,\quad
  \nabla_{\mb}\,\fD_\a\o_{\s\nb}\=\del_{\mb}\,\fD_\a \o_{\s\nb}-\Th_{\mb}{}^\lb{}_\nb \,\fD_\a \o_{\s\lb}~.
\eeq
For the connection to be holomorphic we need to set  $\e-\r{\=}1$. It can be checked that this relation is sufficient to ensure that $\IR^{(0,2)}  = 0$. So we have found a 1-parameter family of holomorphic connections on $\IX$. 

Computing, we find the following non-zero components for the physical deformations~$\fD_\a \Th_\mb$\,:
\beq\label{eq:GammaVariation}
 \begin{split}
  \fD_{\a}\Th^{(\e,\e-1)}{}_\mb{}^\n{}_{\s} 
  &\=\nabla_\s\,\D_{\a\mb}{}^\n+i\,\nabla^{\n}\,\fD_\a\o_{\s\mb}~,\\[0.1cm]
  \fD_{\a}\Th^{(\e,\e-1)}{}_\mb{}^\nb{}_{\sb} 
  &\=\!-g^{\nb\l}\big(\nabla_\l\,\D_{\a\mb}{}^\r+i\,\nabla^\r\fD_\a\o_{\l\mb}\big)\,g_{\r\sb}~.
 \end{split}
\eeq

Before we continue, let us pause to make some comments. Firstly, we have not computed terms which have vertical indices, such as $\fD_\a \Th_\m{}^\a$, as they do not appear in \eqref{eq:modulimetric}.  

Second, it is straightforward to show that $\fD_\a \Th$ satisfies the Atiyah condition:
\beq
\nabla^{(0,1)} \fD_\a \Th^{(0,1)} \= \D_\a{}^\m R_\m~.
\eeq

Third, for the Hull connection $(\e,\r){=}(1,0)$ if we compute the covariant derivative of the fibre metric, we find it vanishes since we have set $S$ to zero:
\beq\notag
 \Inabla_{\a} (\dd s^2_X) \=  \Inabla_{\a} \big(2\,g_{\m\nb} \,e^{(\m} \otimes e^{\nb)}\big) \;=\, -2g_{\m\nb} \big( S_{\a\bb}{}^\m \,\dd y^\bb \otimes e^\nb + S_{\a\bb}{}^\nb \,e^\m \otimes \dd y^\bb \big) \= 0~.
\eeq
These covariant derivatives do not mix components of the fibre metric with components of the base metric under parallel transport along the moduli space. 

Fourth, the extended connection $\ITheta$ defines a covariant derivative of tensors, and it might be tempting to interpret this parallel transport as the appropriate deformation theory of tensors. However, this  does not reduce to known expressions derived in \citeI for the appropriate deformations of tensors on $X$. Note also, if one were to impose that $\Inabla$ and $\pi$ commute, then this would imply $\ITheta^m{}_a$ and $\ITheta^a{}_m$ vanish.  This would mean that $\fD_{\a}g_{\m\nb}$ and $\D_{\a\mb}{}^\n$ both vanish, which is a condition we do not want.

\subsection{The contribution of $\fD_\a \Th$ to the moduli space metric}
\vskip-10pt
We are now in a position to compute the last term of \eqref{eq:modulimetric}. The connection in that metric is the Hull connection $(\e,\r){\=}(1,0)$, though in fact $\e$ drops out of the following calculation and so it is valid for a 1-parameter family.  The integration is evaluated for a fixed point $y\in M$ giving a simplifying rule $e^m \to \dd x^m$.

We use the result \eqref{eq:GammaVariation} to find
\beq\notag
 \begin{split}
  \tr\Big(\fD_\a\Th\star\fD_{\overline\b}\Th\Big)
&=~  2\,\Big(\nabla^{(1,0)}\D_{\a}{}^\m+i\,\nabla^\m\,\fD_\a\o^{(1,1)}\,\Big)\star\Big(\nabla^{(0,1)}\D_{\overline\b}{}^{\nb}-i\,\nabla^{\nb}\,\fD_{\overline\b}\o^{(1,1)}\,\Big)\,g_{\m\nb}~,
 \end{split}
\eeq
where
$\nabla^{(1,0)}\D_{\a}{}^\m{\=}\dd x^\n\,\nabla_\n\,\D_\a{}^\m$. 

Using  $\fD_\a\o^{(0,2)} = \cO(\ap)$ we find
\beq\notag
 \begin{split}
 &-\frac{\ap}{4V}\int_{X}\tr\Big(\fD_\a\Th\star\fD_{\overline\b}\Th\Big)\=\\[0.2cm]
 &\qquad-\frac{\ap}{2V}\int_{X}\nabla^{(1,0)}\D_{\a}{}^\m\star\nabla^{(0,1)}\D_{\overline\b}{}^{\nb}\ g_{\m\nb}\, - \,\frac{\ap}{2V}\int_{X}\nabla^{\m}\,\fD_\a\o\star\nabla_{\m}\,\fD_{\overline\b}\o\\[0.2cm]
 &\qquad+\frac{\ii\ap}{2V}\,\int_{X}\nabla^{(1,0)}\D_{\a}{}^\m\star\nabla_{\m}\,\fD_{\overline\b}\o
\,  - \, \frac{\ii\ap}{2V}\,\int_{X}\nabla_{\nb}\,\fD_\a\o\star\nabla^{(0,1)}\D_{\overline\b}{}^{\nb}+\cO(\ap^2)~.
 \end{split}
\eeq
At this point, we notice a series of useful identities. The variation of the complex structure satisfies
\beq\notag
  \nabla^\s\,\nabla_\s\,\D_\a{}^\m\=
\D_\a{}^{\s\n} R_{\s\lb\n}{}^\m \dd x^\lb~~~\text{and}~~~  
\nabla_\m\nabla^{(1,0)}\D_\a{}^\m \= 0~.
\eeq
where we used the vanishing of the pure part of the curvature tensor for $\Th$: $R_{\n\r\l}{}^\m = 0$. 

For the K\"ahler form variation we find
\beq\notag
 \nabla_\s\nabla^\s\fD_\a\o \= R_{\m\nb}{}^{\s\tb}\,\fD_\a \o_{\s\tb}  \,\dd x^\m \dd x^\nb
~~~\text{and}~~~
  g^{\m\nb}\,\nabla^{\t}\nabla_{\nb}\fD_\a\o_{\t\sb}\,\dd x^{\sb} \= 0~.
\eeq

After integrating by parts, using the terms above and metric compatibility of $\nabla$ we find to first order in $\ap$:
\beq\notag
 -\frac{\ap}{4V}\int_{X}\!\tr\Big(\fD_\a\Th\star\fD_{\overline\b}\Th\Big)\=
\frac{\ap}{2V} \int_X \!\left( \D_{\a\,\mb\nb} \D_{\bb \r\s}+ 
\fD_\a \o_{\r\mb} \,\fD_\bb \o_{\s\nb} \right) R^{\mb\r\nb\s} ~.
\eeq
This expression agrees in form with that derived in \cite{Anguelova:2010ed}.  
\newpage
\section{Deriving the moduli space metric from its \K potential}\label{s:metric}
\vskip-10pt
We derive the moduli space metric from the \K potential in a concise manner using extended forms on $\IX$. As in the previous section, we set $S_{ab} = 0$ and within integrals over $X$ we have the rule $e^m \to \dd x^m, e_a \to \del_a$. 
 
 The moduli space metric $g^\sharp_{\a\bb}$ has the associated \K form
$$
\o^\sharp \= \ii g^\sharp_{\a\bb}\, \dd y^\a \dd y^\bb~.
$$
We show using $\IX$ that 
$$
 \o^\sharp \= \ii\, \fD \fDb \ccK  ~,~~~\text{where}~~~ 
\ccK \= \ccK_1 + \ccK_2 ~= - \log \left( \frac{4}{3} \int\!\o^3  \right) - \log\left( \ii  \int\!  \O\,\Ob \right)~.
$$
That is, we are showing that $\ccK$ is the \K potential for the moduli space metric. 

We adopt the convention that when a universal form appears within an integral over $X$, the only surviving part is that which makes the integrand a top form on $X$. Some useful statements illustrating this are
\beq
\begin{split}
 \int_X\!\o^2 \IF 
&\= \int_X\! \o^2 F \= 0~, \\[8pt]
\half\! \int_X\! \o^2 \tr \IF^2 
&\=  \int_X\! \o^2 \tr \big(F\, \IF_{\a\bb} -\IF_\a\, \IF_\bb \big)\, \dd y^\a \dd y^\bb \= 
\int_X\! \o^2 \tr (\fD_\a \A\, \fD_\bb \A^\dag)\, \dd y^\a \dd y^\bb ,\\[8pt]
\int_X \!\!\o^2\, \fD\,\fDb\, \o 
&\= \int_X\! \o^2 \Idel \Idelb\hskip1.5pt \o \= \int_X\! \o^2 \Idel \Idelb \Iomega~.\raisetag{1cm}
\end{split}\label{useful1}\eeq
where  we use the relations $\o^2 F {\=} 0$, $\fD_\bb \A {\=} 0$ and $\delb(\o^2) {\=}\del(\o^2) {\=} 0$. We will also use, within the integrand,
\vskip-20pt
\beq
\fD \left( \frac{\o^2}{2V} \right) ~= -\frac{1}{V} \star \fD \o~.
\label{useful2}\eeq
Recall that $\Id \O \;= - k^\sharp \O + \Ichi$  with $\Ichi \= \half \chi_{\a \m\n\rb}\, \dd y^\a \dd x^\m \dd x^\n \dd x^\rb$ and
$k^\sharp {\=}\fD \ccK_2= \dd y^\a \del_\a \ccK_2$ and we use $\dd \O {\=} \fDb \O {\=} 0$.  

Consider first the derivatives of $\ccK_1$,
\beq
\begin{split}
\,  \fD \fDb \ccK_1 &\=\! - \fD \left( \frac{1}{2V} \int_X \o^2 \fDb \o \right)
\\[6pt]
 &\= \+ \frac{1}{V} \int_X \fD \o \star \fDb \o  - \frac{\ii}{2V} \int_X \o^2 \Idel\Idelb \Iomega~
\\[6pt]
 &\= \+ \frac{1}{V} \int_X \fD \o \star \fDb \o - \frac{\ap}{16V} \int_X \o^2 
 \left(  \tr \IF^2 -  \tr \IR^2 \right) ~
\\[6pt]
 &\= \left( \frac{1}{V} \int_X \fD_\a \o \star \fD_\bb \o + \frac{\ii\ap}{8V} \int_X \o^2  \tr\! \left( \fD_\a \A \fD_\bb \A^\dag - \fD_\a \th \fD_\bb \th^\dag   \right) \right) \dd y^\a \dd y^\bb ~,
\end{split}\label{eq:om1}\raisetag{3.2cm}
\eeq
\nobreak where we have used \eqref{useful1} and \eqref{useful2}. \goodbreak

While for the derivatives of $\ccK_2$ we have
\beq
\begin{split}
 \ii \fD \fDb \ccK_2 ~
 &\=\! -\ii \fD \left( \frac{\int_X \O\, \fDb \Ob }{\int_X \O\, \Ob} \right)
\\[8pt]
 &\= \+ \ii \left( \frac{ \int_X \fD \O \, \Ob \; \int_X \O\, \fDb \Ob}{(\int_X\O\,\Ob)^2} -
  \frac{\int_X \fD \O\, \fDb \Ob}{\int_X \O\, \Ob}\right)
\\[8pt]
 &\= - \ii \, \frac{\int_X \chi_\a\, \chib_\bb}{\int_X \O\, \Ob} \;\dd y^\a \dd y^\bb~,
\end{split}\label{eq:om2}
\eeq
where we use $\Idelb \O {\=} \delb\O + \fDb \O {\=} 0$ and, in the second line, several terms vanish owing to considerations of holomorphic type. 

Finally, combining \eqref{eq:om1} and \eqref{eq:om2}, we obtain the desired result
\beq\notag
  \ii\,  \fD \fDb K \=\o^\sharp~.
\eeq
\vfill
\section*{Acknowledgements}
\vskip-10pt
We gratefully acknowledge interesting conversations with Anthony~Ashmore, Marc-Antoine Fiset, Nigel~Hitchin, Richard~Thomas, Eirik~Svanes and Dan~Waldram. PC, JM and XD wish to  acknowledge also the hospitality IISc Bangalore and the String theory, the Mainz Institute for Theoretical Physics where part of this work was done. PC and XD also wish to thank KIAS for hospitality while this work was completed and have been supported by EPSRC grant BKRWDM00. JM~is supported by STFC grant ST/L000490/1. 
\newpage
\appendix
\section{Heterotic Geometry}
\label{s:HetGeometry}
\vskip-10pt
The purpose of this section is to establish the notation that we use in this paper, and to review some material derived in \citeI and \cite{Eguchi:1980jx}. 

\subsection{Some differential geometry}
\vskip-10pt
This material is standard, and used  in section 4. Our notation follows that of \cite{Eguchi:1980jx,Becker:2007zj}, and more detail may be found in those references.

Consider a  manifold $X$ with metric given by
$$
\dd s^2 \= g_{mn}\, \dd x^m \otimes \dd x^n \= 2g_{\m\nb} \,\dd x^\m \otimes \dd x^\nb.
$$
We introduce a basis of orthonormal 1-forms $s^A = s^A{}_m \dd x^m$ so that 
$\dd s^2 {\=} \d_{AB} s^A \otimes s^B$ 
and an affine spin connection 1-form $\Th$ which satisfies
\beq\label{eq:torsion}
\dd s^A + \Th^A{}_B \wedge s^B = \half T^A{}_{BC} \,s^B \w s^C \= \half T^A{}_{mn} \dd x^m \dd x^n, \qquad \Th^A{}_B \= \Th_m{}^A{}_{B}\,  \dd x^m~,
\eeq
where $T^A{}_{[BC]}$ is the torsion. 
 In this section  only we use upper case roman letters $A,B,\cdots$ to denote flat Lorentz indices. The ordering of the indices here is important, and without care, can lead to sign errors. In the following, we omit the wedge symbol `$\wedge$' where possible.

The connection $\Th$ being metric compatible means that $\Th_{AB} = - \Th_{BA}$, where the indices are raised and lowered with the flat metric $\d_{AB}$. The Levi--Civita connection $\Th^{\LC}$ is the unique connection that is both metric compatible and has vanishing torsion. The Levi--Civita connection has symbols, in a coordinate basis, given by the following transformation law
\beq\label{eq:ThLC}
\Th^{\LC}{}_m{}^A{}_{B} \= s^A{}_n\del_m S_B{}^n + s^A{}_q\G^{\LC}{}_{p}{}^q{}_m  S_B{}^p~.
\eeq

where $S_A = S_A{}^p \del_p$ is a basis of vectors dual to the 1-forms $s^A$. In the coordinate basis $\G^{\LC}$ is symmetric in its lower indices; in other bases this is not necessarily the case.  In the literature, the $\G$ symbol is often written with  a change of ordering of the indices $\G_{mB}{}^A = \Th_m{}^A{}_{B} $. Care must be taken with signs and for this reason we largely avoid writing the $\G$ symbols. 
In the coordinate basis, the symbols of Levi--Civita connection are the Christoffel symbols 
\beq\label{eq:GLC}
\G^{\LC}{}_m{}^p{}_q \= \half g^{pl} \left( \del_m g_{lq} + \del_q g_{lm} - \del_l g_{mq} \right)~.
\eeq

The curvature 2-form is defined as
$$
R^A{}_B \= \dd \Th^A{}_B + \Th^A{}_C\, \Th^C{}_B \= \half R^A{}_{BCD}\, s^C \,s^D \= \half R^A{}_{Bmn} \dd x^m \dd x^n~.
$$
We will have a need also for spinors. We denote the hermitian gamma matrices on $X$ by~$\g^m$; these satisfy the Clifford algebra $\{ \g^m , \g^n \} = 2 g^{mn}$. The anti-symmetrised product of $k$ gamma matrices is 
$$
\g^{m_1\cdots m_k} \= \g^{[m_1} \cdots \g^{m_k]} ~, 
$$
 and the covariant derivative acting on a spinor $\ve$ on $X$ is 
\beq
\nabla_m \ve \= \del_m \ve + \frac{1}{4} \Th_{mAB} \g^{AB} \ve~.
\eeq
Suppose $\ve$ is a  non--vanishing Weyl spinor on $X$ and is covariantly constant $\nabla_m \ve = 0$. We normalise so that $\ve^\dag \ve = 1$. We can then use $\ve$ to define tensors as spinor bilinears. 
The complex structure $J = J_m{}^n\, \dd x^m \otimes \del_n = J_A{}^B \,s^A \otimes S_B$ will also be relevant. As a spinor bilinear it is
$$
J_m{}^n \= -i \ve^\dag \g_{m}{}^n \ve~.
$$
It is covariantly constant
\beq\label{eq:nablaJ}
\nabla_m J \= \left(\del_m J_A{}^B +  \Th_m{}^A{}_C\,J_B{}^C -  \Th_m{}^C{}_B \,J_C{}^A\right) s^A \otimes S_B \= 0~.
\eeq
It can be shown that the Nijenhuis tensor of $J$ vanishes \cite{Strominger:1986uh}, so the manifold is  complex. We denote its complex coordinates $x^\m, x^\nb$, then~$J_\m{}^\n {=} \ii \d_\m{}^\n$ and~\hbox{$J_\mb{}^\nb {=} - \ii \d_\mb{}^\nb$}. The  compatibility equation between the hermitian form $\o_{mn}$, metric $g_{mn}$ and complex structure $J_m{}^n$ is
\beq
\o_{mn} \= J_m{}^p g_{pn}~.
\eeq
In complex coordinates $\o_{\m\nb} = \ii g_{\m\nb}$.

\subsection{Heterotic action and supersymmetry variations}
\vskip-10pt
The supergravity action, correct to up to and including $\ap^2$ is the following
 \cite{Bergshoeff:1989de,Bergshoeff:1988nn}:
\begin{equation}
S = \frac{1}{2\kappa_{10}^2} \int\! \dd^{10\,}\! X \sqrt{g_{10}}\, e^{-2\Phi} \Big\{ \cR -
\half |H|^2  + 4(\del \Phi)^2 - \frac{\alpha'}{4}\big( \tr |F|^2 {-} \tr |R(\Th^\Hu)|^2 \big) \Big\} + \cO(\alpha'^3),
\label{eq:10daction}
\end{equation}
 The 10D Newton constant is denoted by $\kappa_{10}$,
\hbox{$g_{10}=-\det(g_{MN})$}, $\Phi$ is the 10D dilaton and $\cR$ is the Ricci scalar evaluated using the Levi-Civita connection. 

The point-wise inner product on $p$-forms is
$$
|T|^2\= 
\frac{1}{p!} \, g^{M_1 N_1} \ldots g^{M_p N_p}\, T_{M_1\ldots M_p} \,T_{N_1 \ldots N_p}~.
$$
Thus the curvature squared terms correspond to
$$
\tr |F|^2 = \half \tr F_{MN} F^{MN}~~~\text{and}~~~ \tr |R(\Th^\Hu)|^2 \=
\half \tr R_{MNPQ}(\Theta^\Hu) R^{MNPQ}(\Theta^\Hu)~,
$$
where the Riemann curvature is evaluated using 
$$
\Th^\Hu{} \= \Th^{\LC} + \half H~. 
$$
Using \eqref{eq:torsion} this connection has torsion given by  
$$
T^A{}_{MN} \=\! -  H^A{}_{MN} ~.
$$
The connection $\nabla^\Hu$ has torsion with the opposite sign to the connection $\nabla^\Bi$ which appears in the supersymmetry variations. In string frame, these are
\beq
\begin{split}
\d \Psi_M & \= \nabla^\Bi \ve \= \nabla^{\LC}_M \ve - \frac{1}{4} H_M \ve \= 0~,\\[3pt]
\d \lambda &\= - \half (\delslash \Phi )\ve + \frac{1}{4} \Hslash \ve \= 0~, \\[3pt]
\d \chi &\= - \half \Fslash \ve \= 0~,
\end{split}
\eeq
corresponding to the gravitino, dilatino and gaugino variations. We have introduced 
$$
H_M \= \half  H_{MNP} \G^{NP}~, \quad \Hslash \= \frac{1}{3!} H_{MNP} \G^{MNP}~, \quad \Fslash \= \half F_{MN}\G^{MN}~, \quad \delslash \= \G^M \del_M~.
$$
 We will always assume the dilaton is constant, and work in weakly coupled perturbative string theory at large radius. The spacetime geometry is $\IR^{3,1}\times X$ where the manifold $X$ is compact. The gamma matrices $\G^M$ and spinor $\ve$ are decomposed in a manner compatible with this direct product 
$$
\G^e \= \g^e \otimes 1~, \qquad \G^m \= \g^5 \otimes \g^m~, \quad \ve \= \zeta_+ \otimes \eta_+ + \zeta_- \otimes \eta_-~,
$$
where $e,f,\cdots$ are spacetime indices, $\zeta_\pm$ are Weyl spinors on $\IR^{3,1}$ and $\eta_\pm$ are Weyl spinors on $X$ with $\zeta_+^* = \zeta_-$ and \hbox{$\eta_+^* {\=} \eta_-$}. 

From the calculation in the previous subsection, equation \eqref{eq:nablaJ} requires $J_m{}^n$ to be covariantly constant with respect to $\nabla^\Bi$:
\beq\label{eq:Jcovconstant}
\nabla_m^\Bi J_n{}^p \= \nabla^{\LC}_m J_n{}^p - \frac{1}{2} H_m{}^p{}_q \,J_n{}^q + \frac{1}{2} H_m{}^q{}_n \,J_q{}^p \= 0~.
\eeq
Contracting with  $g_{pr} J_s{}^r \dd x^m \dd x^n\dd x^s$, gives 
$$
  \dd \o +\half H_{mnp} \dd x^m \dd x^n J^p \= 0~.
$$
Evaluating this equation in  complex coordinates gives 
\beq\label{eq:Jcovconstant2}
\ii (\del - \delb) \o \= H~.
\eeq

 The manifold $X$ admits a holomorphic volume form $\O$ which  is related to its hermitian form $\o$ through a compatibility relation
\begin{equation}
\frac{1}{3!} \o^3 \=\frac{\ii\,\O\wedge\Ob}{\norm{\O}^2}~.
\end{equation}
The holomorphic  form $\O$ is $\delb$-closed and satisfies $\star \O = - \ii \O$, which also means it is $\delb$-harmonic. Similarly,  $\O$ is covariantly constant with respect to $\Th^\Bi$. 
This implies that
$$
H_{\m\n}{}^\n \= 0~~~\text{and}~~~ \del_\m \log \norm{\O}^2 \= 0~.
$$

\subsection{The Background field expansion}
\vskip-10pt
The background field expansion is a small fluctuation expansion around a classical background, in which the small fluctuations modulo gauge redundancies are the dynamical quantities of physical interest. We describe this for heterotic theories following the background field method.  

Consider a small fluctuation $A\to A{+}\d A$, together with a variation of the gauge transformation $\Phi\to\Phi(1+\e)$. 
The total quantity $A+\d A$ transforms by \eqref{eq:Atransf}. The doctrine of the background field method assigns $A$ the transformation law \eqref{eq:Atransf} while the fluctuation transforms as 
$$
\d A~\to~ \Ph\big( \d A - \dd_A\e\big) \Ph^{-1}~.
$$
This  is understood as the  composition of two gauge transformations:
\begin{itemize}
\item Background gauge transformations
\beq\label{eq:BackgroundA}
A~\to~ \Ph A \Ph^{-1} -\, \dd\Ph\,\Ph^{-1}~~~\text{and}~~~
\d A~\to~ \Ph\, \d A\, \Ph^{-1}~.
\eeq

\item Small gauge transformations
\beq\label{eq:smallGauge}
A~\to~A~~~\text{and}~~~\d A~\to~\d A -\dd_A\e~.
\eeq
\end{itemize}
The former is a classical symmetry of the background, while the latter describes the gauge redundancies of the dynamical variables. Although we have described this using deformations $\d A$ it equally well applies to other gauge symmetries such as Lorentz or diffeomorphisms. The former acts on all tensors, including $\d \Th$ and the metric $\d g$. The B-field $\d B$ has an additional symmetry through its Gerbe property.

To first order, variations of heterotic structures  are related by differentiating $H = \dd^c \o$ and  using the cohomology of $X$: 
\beq
\begin{split}
\ccB_\a^{(2,0)}  &\= \del \b_\a^{1,0}~,\\[8pt] 
\fD_\a\o^{(2,0)} &\= 0~,\\[10pt]
\ccB_\a^{(0,2)} + \ii \fD_\a \o^{(0,2)} &\= \delb \k_\a^{0,1}~, \\[3pt] 
\ccB_{\a}^{(1,1)} - \ii \fD_\a \o^{(1,1)} &\= \Big(\g_\a + \dd\left( \a^{0,1}_\a + \b^{1,0}_\a\right)\Big)^{(1,1)}~,\\[3pt]
\delb (\ccB_\a^{(1,1)} + \ii \fD_\a\o^{(1,1)} - \del \k_\a^{(0,1)} ) &\= 2\ii \D_\a{}^\m (\del_\m \o - \del \o_\m) + \frac{\ap}{2} \tr( \fD_\a \A F) - \frac{\ap}{2} \tr( \fD_\a \th R)~, \raisetag{3cm}
\end{split}\label{eq:cBrelations2}
\eeq
where $\g_\a^{(1,1)}$ is $\dd$-closed $(1,1)$-form. As $\ccB_\a$ is defined up to $\dd$-closed form, we can absorb the terms involving $\b_\a^{(1,0)}$ and $\a_\a^{(0,1)}$. We show in \citeI that  $\g_\a$ can be absorbed by an $\ap$-correction to the moduli space coordinates.   

Using the covariant derivatives of fields as a basis for a Kaluza--Klein reduction, with the harmonic gauge fixing, gives the moduli space metric \eqref{eq:modulimetric}. It is \K  after taking into account the second order relations between fields. This observation can be generalised to account for the charged matter fields and their fermionic superpartners in order to give the matter field metric as derived in \cite{McOrist:2016cfl}. This  normalises physical Yukawa~couplings.

\section{The $\IGamma$ and $\ITheta$ symbols}
\label{s:CovDerivC}
\vskip-10pt
The calculation in this section follows that of \cite{YanoBook}. The two relevant bases for $\ccT_{\IX}$ are the coordinate basis and the $e$-basis
$$
 \del_P\=\{\del_a,\,\del_m\}~ , \qquad \qquad e_P\=\{e_a,\,e_m\}\=\{\del_a-c_a{}^m\,\del_m,\,\del_m\}\ ,
$$
While for $\ccT^*_{\IX}$ they are
$$
 \dd u^P\=\{\dd y^a,\,\dd x^m\}~ ,\qquad \qquad e^P\=\{e^a,\,e^m\}\=\{\dd y^a,\,\dd x^m+c_a{}^m\,\dd y^a\}~.
$$
We introduce  the matrix $e_Q{}^P$ and its inverse $E^P{}_Q$ as follows
\beq\notag
e^P\= e_Q{}^P\,\dd u^Q~, \qquad  e_Q\= E^P{}_Q\,\del_P~.
\eeq
More explicitly
\beq
\begin{split}
 e_Q{}^P &\=
 \begin{pmatrix}
  \d_b{}^a & 0 \\[0.2cm]
  c_b{}^m & \d_n{}^m 
 \end{pmatrix}\ ~,\qquad
 E^P{}_Q\=
 \begin{pmatrix}
  \d^a{}_b & 0\\[0.2cm]
  -c_b{}^m & \d^m{}_n
 \end{pmatrix}~.
 \end{split}\label{eq:changeofbasis}
\eeq
The covariant derivative $\Inabla$ defines symbols  $\ITheta$ in the $e$-basis and $\IGamma$ in the coordinate basis:
\beq\notag
 \Inabla(\del_Q)\=\dd x^M\,\IGamma_M{}^P{}_Q\,\del_P~~~\text{and}~~~ \Inabla(e_Q)\= e^M\,\ITheta_M{}^P{}_Q\,e_P~.
\eeq
The relation between the symbols follows from
\beq\notag
\Inabla(e_Q) \= e^M\Big( e_M (E^P{}_Q)  + E^N{}_ME^S{}_Q \IGamma_N{}^P{}_S \,\Big)\del_P \= e^M \ITheta_M{}^S{}_Q\, E^P{}_S\, \del_P~.
\eeq
which we rewrite  as
\beq\label{eq:Yano}
 e_M ( E^P{}_Q)  + E^N{}_ME^S{}_Q \IGamma_N{}^P{}_S   \= \ITheta_M{}^S{}_Q\, E^P{}_S ~.
\eeq

Under a coordinate transformation, $x \to \wt x(x,y)$ and $y \to \wt y(y)$, the shift $c_a{}^m$ transforms  so that the $e$-bases elements rotate in a block diagonal fashion 
$$
e_a \to \wt e_a \= j_a{}^b \,e_b~, \qquad \del_m \to \wt \del_m \=  j_m{}^n \del_n~, \qquad j_a{}^b \= \frac{\del y^b}{\del \wt y^a}~, \qquad j_n{}^m \= \frac{\del  x^n}{\del \wt x^m}~.
$$
This is viewed as a block diagonal rotation of the $e$-basis 
$$
j_P{}^Q  \= \begin{pmatrix} j_a{}^b & 0 \\ 0 & j_m{}^n \end{pmatrix}~.
$$
The $\ITheta$ symbols therefore transform as 
$$
\wt \ITheta_M{}^P{}_Q \= \Big(\wt e_M \left( j_Q{}^S\right) + j_M{}^T j_Q{}^N \ITheta_T{}^S{}_N\Big) j^{-1}{}^P{}_S~.
$$
The block diagonal structure ensures that symbols such as $\ITheta_m{}^n{}_a$ and $\ITheta_m{}^a{}_n$ transform as tensors, and so their geometric meaning is independent of our choice of $c$:
\beq\notag
\begin{split}
\wt \ITheta_m{}^n{}_a &\= j_m{}^p j^{-1}{}^n{}_q \,j_a{}^b \ITheta_p{}^q{}_b~,\qquad \wt \ITheta_m{}^a{}_n \= j_m{}^p j{}_n{}^q j^{-1}{}^a{}_b\, \ITheta_p{}^b{}_q~.\\
\end{split}
\eeq
The geometric interpretation of these  symbols is that they are the components of the extrinsic curvature $\chi_{am}{}^n$ of the fibration of $X$ in $\IX$ as described in section \sref{s:xtrinsic}.

The  covariant derivative for $c_a{}^m$ follows by choosing $M{\=}m$, $P{\=}p$, $Q{\=}a$ in \eqref{eq:Yano} and using \eqref{eq:changeofbasis} together with $\ITheta_m{}^b{}_a {\=} \IGamma_m{}^b{}_a {-} \IGamma_m{}^b{}_q c_a{}^q$:
\beq\label{eq:CovDerivC}
 \chi_{am}{}^p\= \ITheta_m{}^p{}_a \=\! -( \del_m c_a{}^p +   \IGamma_m{}^p{}_n c_a{}^n -  \IGamma_m{}^p{}_a +   \IGamma_m{}^b{}_a c_b{}^p - \IGamma_m{}^b{}_q c_a{}^q c_b{}^p) ~.
\eeq
The extrinsic curvature $\chi$ is a tensor and so also is the final expression above.  We could, following \cite{YanoBook}, regard this final expression as the definition of a  new covariant derivative of the shift $c_a{}^m$. 

When $S_{ab} = 0$, the expression simplifies 
\beq\notag
 \chi_{am}{}^p \=\!  -(\del_m c_a{}^p +   \IGamma_m{}^p{}_n c_a{}^n -  \IGamma_m{}^p{}_a)~.
\eeq

We can make some comments on this equation.
\begin{enumerate}
 \item If $c_a{}^m = 0$, so that normal vectors are given by $\del_a$, then
  $$
\chi_{am}{}^p \= \IGamma_m{}^p{}_a ~.
 $$
 The following symbols are trivially identical $\ITheta_m{}^p{}_a = \IGamma_m{}^p{}_a$. 
 
 \item In complex coordinates if we choose the Levi--Civita connection for $\Inabla$ and use  the  $\IGamma$ symbols of  \eqref{eq:GammaSymbols}, we find
 $$
 \chi_{\a\mb\rb} \= \D_{\a(\mb\rb)}~.
 $$

\item An analogous calculation gives
$$
\chi_{\a\mb\r} \= \half \fD_\a g_{\r\mb}~.
$$

\end{enumerate}

\vskip10pt
\subsection{The $\IGamma$ Symbols for the Levi--Civita connection}
\label{app:LeviCivita}
\vskip-10pt
These are the $\IGamma$ symbols  for the Levi--Civita connection. 
We first invert the relation \eqref{eq:Yano} and decompose the indices, giving
\beq\label{IThtoIG}
\begin{split}
  &\IGamma^m{}_n\=\!-c_a{}^m\,\ITheta^b{}_n+\ITheta^m{}_n\ ,\\[0.1cm]
  &\IGamma^a{}_n\=\ITheta^a{}_n\ ,\\[0.1cm]
  &\IGamma^m{}_b\=\!-c_a{}^m\,\ITheta^a{}_b-c_a{}^m\,\ITheta^a{}_n\,c_b{}^n+\ITheta^m{}_b+\ITheta^m{}_n\,c_b{}^n+\Id\,c_b{}^m\ ,\\[0.1cm]
  &\IGamma^a{}_b\=\ITheta^a{}_b+\ITheta^a{}_n\,c_b{}^n~.\\[0.1cm]
 \end{split}
\eeq
Using the symbols for $\ITheta^\LC$ from \eqref{eq:LConIX}, we have
\beq
\begin{split}
  &\IGamma^n{}_k\=\dd x^m\Big(\G^{\LC}{}_m{}^n{}_k + \frac{1}{2}\,c_b{}^n\,g\#{}^{bd}\,\fD_d g_{mk}\Big) + \\[0.1cm]
  &\hspace{1.5cm} + \dd y^a\Big(c_a{}^m\,\G^{\LC}{}_m{}^n{}_k + c_a{}^m\,c_b{}^n\,g\#{}^{bd}\,\fD_d g_{mk}  +  \del_k c_a{}^n + \frac{1}{2}\,g^{nl}\,\fD_a g_{lk}  +  c_b{}^n\,g\#{}^{bd}\,S_{ad}{}^l\,g_{lk}\Big)\ ,\\[0.3cm]
    &\IGamma^b{}_k\=\!-\frac{1}{2}\,\dd x^m\,g\#{}^{bd}\,\fD_d g_{mk} - \frac{1}{2}\dd y^a\Big(c_a{}^m\,g\#{}^{bd}\,\fD_d g_{mk} + g\#{}^{bd}\,S_{ad}{}^l\,g_{lk}\Big)\ ,\\[0.3cm]
  &\IGamma^n{}_c\=\dd x^m\Big(\G^{\LC}{}_m{}^n{}_k\,c_c{}^k + \del_m c_c{}^k  + \frac{1}{2}\,c_b{}^n\,g\#{}^{bd}\,\fD_{d} g_{mk}\,c_c{}^k + \frac{1}{2}\,g^{nl}\,\fD_c g_{lm} +\frac{1}{2}\,c_b{}^n\,g\#{}^{bd}\,S_{cd}{}^l\,g_{lm}\Big)\\[0.1cm]
  &\hspace{1.5cm} + \dd y^a\Big( \del_a c_c{}^n + (\del_k c_a{}^n)c_c{}^k + c_a{}^m\,c_b{}^n\,c_c{}^k\G^{\LC}{}_m{}^n{}_k - c_b{}^n\,\G\#{}^{\LC}{}_a{}^b{}_c \\[0.1cm]
  &\hspace{3cm}  + \frac{1}{2}\,c_a{}^m\,c_b{}^n\,g\#{}^{bd}\,\fD_d g_{mk}\,c_c{}^k + \frac{1}{2}\,c_a{}^m\,g^{nl}\,\fD_c g_{lm} + \frac{1}{2}\,c_c{}^k\,g^{nl}\,\fD_a g_{lk}  \\[0.1cm]
  &\hspace{3.5cm} + \frac{1}{2}\,S_{ac}{}^n + \frac{1}{2}\,S_{cd}{}^l\,c_a{}^m\,c_b{}^n\,g\#{}^{bd}\,g_{lm} + \frac{1}{2}\,S_{ad}{}^l\,c_b{}^n\,c_c{}^k\,g\#{}^{bd}\,g_{lk}\Big)\ ,\\[0.3cm]
  &\IGamma^b{}_c\=\!-\frac{1}{2}\,\dd x^m\Big(c_c{}^k\,g\#{}^{bd}\,\fD_d \,g_{mk}  - S_{dc}{}^l\,g\#{}^{bd}\,g_{lm} \Big)\\[0.1cm]
  &\hspace{1.5cm} + \dd y^a\Big(\G\#{}^{\LC}{}_a{}^b{}_c - \frac{1}{2}\,c_a{}^m\,c_c{}^k\,g\#{}^{bd}\,\fD_d g_{mk} + \frac{1}{2}\,S_{dc}{}^l\,c_a{}^m\,g\#{}^{bd}\,g_{lm}  - \frac{1}{2}\,S_{ad}{}^l\,g\#{}^{bd}\,g_{lk}\,c_c{}^k\Big)\ .
 \end{split}\label{eq:GammaSymbols}
 \raisetag{5.2cm}
\eeq

\section{The Nijenhuis tensor for $\IX$}\label{app:Nij}
\vskip-10pt
The Nijenhuis tensor for $\IJ$ is
\beq
N_\IJ \=( \IJ^P \del_P \IJ^Q - \IJ_P{}^Q \Id \IJ^P)\del_Q~,\label{eq:nijIJ}
\eeq
where $u^P = (y^a,x^m)$ denotes a point in $\IX$ and we write $\IJ^P = \IJ_S{}^P \dd u^S$. 
The complex structure is triangular in the coordinate basis: 
$$
\IJ\= J_m{}^n e^m \otimes e_n + J^\sharp{}_a{}^b e^a \otimes e_b \=  J_m{}^n\, \dd x^m \otimes \del_n + ( c_a{}^m J_m{}^n - J^\sharp{}_a{}^b c_b{}^n) \,\dd y^a \otimes \del_n + J^\sharp{}_a{}^b \,\dd y^a \otimes \del_b~. 
$$
Thus, 
$$
\IJ_m{}^a \= 0, \qquad \IJ_a{}^m = J_n{}^m c_a{}^m - J^\sharp{}_a{}^b c_b{}^m~.
$$
The terms in \eqref{eq:nijIJ} decompose according to tangibility. In the following, we  suppress the $\otimes$ in writing out the tensor structure of $N_\IJ$ to simplify notation, so for example $N_\IJ = \half N_\IJ{\,}_{PQ}{}^R \dd u^P \dd u^Q \del_R$.

\begin{enumerate}
 \item The first term, proportional to $\dd x^m \dd x^n$, reduces to that on $X$
\beq\label{eq:nij_term1}
\half N_\IJ{\,}_{mn}{}^Q \dd x^m \dd x^n\del_Q \=  N_J~.
\eeq

\item The next term has mixed tangibility $\dd y^a \dd x^m$
\beq\begin{split}
 N_\IJ{\,}_{am}{}^Q \dd y^a \dd x^m\del_Q &\=   N_J{\,}_{mn}{}^q c_a{}^m\, \dd y^a \dd x^n \,\del_q+(\Jsh_a{}^b \d_p{}^q - \d_a{}^b J_p{}^q) \,e_b ( J_m{}^p) \,\del_q \,+ \\[2pt]
 &\hskip30pt +\Big(\Jsh_a{}^b J_p{}^q \d_m{}^n - \Jsh_b{}^a J_m{}^n \d_p{}^q  \Big) [e_n, e_b]^p\,\dd y^a \dd x^m \del_q~.
\end{split}\notag\eeq
where
$$
e_b ( J_m{}^p) \= \del_b J_m{}^p - c_b{}^n \del_n J_m{}^p~~~\text{and}~~~[e_n, e_b] \=\! -(\del_n c_b{}^q)\,\del_q~.
$$
We use the projectors of \eqref{eq:projectors} to rewrite the $N_\IJ{\,}_{am}{}^Q$ components
\beq\begin{split}
 N_\IJ{\,}_{am}{}^Q \dd y^a \dd x^m\del_Q &\= N_J{\,}_{mn}{}^q c_a{}^m\, \dd y^a \dd x^n \,\del_q +  
 2\ii\, (P_a{}^c Q_p{}^q - Q_a{}^c P_p{}^q) e_c ( J_m{}^p)\,  \del_q \,+ \\[2pt]
 &\hskip30pt +4 \Big(P_a{}^c P_m{}^n Q_p{}^q + Q_a{}^c Q_m{}^n P_p{}^q  \Big) [e_n, e_c]^p\, \dd y^a \dd x^m \del_q~.
\end{split}\label{eq:nij_term2}\raisetag{20pt}
\eeq

\item The final term of \eqref{eq:nijIJ} has tangibility $\dd y^a \dd y^b$

\begin{align}\notag
\half &N_\IJ{\,}_{ab}{}^Q \dd y^a \dd y^b \del_Q \=  \half N_\Jsh{\,}_{ab}{}^d \dd y^a \dd y^b\, e_d  + 
\half N_J{\,}_{mn}{}^q c_a{}^m c_b{}^n\, \dd y^a \dd y^b \,e_q\, +\notag\\[2pt]
&  \Big(\Jsh_a{}^c \,e_c (J_p{}^q) c_b{}^p - J_m{}^q \,e_a  (J_p{}^m) c_b{}^p\Big) \dd y^a \dd y^b \del_q\, + \notag\\
& \Big( \d_a{}^c \d_b{}^d \d_p{}^q + \Jsh_a{}^c J_p{}^q \d_b{}^d - \Jsh_a{}^c \Jsh_b{}^d \d_p{}^q +\Jsh_b{}^d J_p{}^q \d_a{}^c\Big) (\del_c c_d{}^p) \dd y^a \dd y^b \del_q\, + \notag\\[2pt]
& \Big( \d_a{}^c \d_b{}^d J_n{}^m J_p{}^q {-} 
\d_a{}^c \Jsh_b{}^d J_n{}^m \d_p{}^q {-} 
\Jsh_a{}^c \d_b{}^d \d_n{}^m J_p{}^q {+} 
\Jsh_a{}^c \Jsh_b{}^d \d_n{}^m \d_p{}^q \Big) c_c{}^n (\del_m c_d{}^p)\, \dd y^a \dd y^b \,\del_q\;.\notag\\
\intertext{In terms of projectors}
\half &N_\IJ{\,}_{ab}{}^Q \dd y^a \dd y^b \del_Q \=  \half N_\Jsh{\,}_{ab}{}^d \dd y^a \dd y^b\, e_d  + 
\half N_J{\,}_{mn}{}^q c_a{}^m c_b{}^n\, \dd y^a \dd y^b \,e_q\, +\notag\\[2pt]
& 2\ii \Big( P_a{}^c Q_p{}^q - Q_a{}^c P_p{}^q \Big) 
\,e_c ( J_m{}^p) c_b{}^m \dd y^a \dd y^b \del_q\, -  
2 \Big( P^c P^d Q_p + Q^c Q^d P_p \Big)  [e_c, e_d]^p\,- \notag\\[2pt]
&4 \Big( P^c P^d Q_p P_n{}^m + Q^c Q^d P_p Q_n{}^m + P^c Q^d Q_n{}^m P_p + 
Q^c P^d P_n{}^m Q_p  \Big) c_c{}^n [e_m, e_d]^p\;.
\label{eq:nij_term3} \end{align}
\end{enumerate}

Gathering \eqref{eq:nij_term1}, \eqref{eq:nij_term2} and \eqref{eq:nij_term3}, and simplifying we find 
\beq\begin{split}
 &N_\IJ \= \half N_J{\,}_{mn}{}^q e^m e^n e_q + 
 2\ii \,e_c\,( J_m{}^p )\, \Big( P^\sharp{}^c e^m Q_p - Q^\sharp{}^c e^m P_p \Big)\, - \\[2pt]
&4 [e_c, e_n]^p \Big(P^\sharp{}^c P^n Q_p + Q^\sharp{}^c Q^n P_p \Big) -2 [e_c, e_d]^p \Big(P^\sharp{}^c P^\sharp{}^d Q_p + Q^\sharp{}^c Q^\sharp{}^d P_p \Big) + 
\half N_\Jsh{\,}_{ab}{}^c e^a e^b e_c\;.
\end{split}\label{eq:nij_term4}\raisetag{50pt}\eeq
The second and third terms  combine  in virtue of the relation
$$
[P^\sharp_a, P_m]^p \dd y^a e^m \= - \frac{\ii}{2} P^\sharp{}^c \,e_c( J_m{}^p) e^m + P^\sharp{}^c P^m [e_c, e_m]^p~.
$$ 
Note also that
$$
[P^\sharp_c , P^\sharp_d]^q \= P^\sharp{}_c{}^a P^\sharp{}_d{}^b [e_a, e_b]^q~. 
$$
These relations, together with \eqref{eq:nij_term4}, give the final expression 
\beq\notag
\begin{split}
  N_\IJ ~&=~ \half N_J{\,}_{mn}{}^q \,e^m \,e^n \,e_q - 4 [P^\sharp_a, P_m]^q \,e^a \,e^m\, Q_q - 4[Q^\sharp_a, Q_m]^q \,e^a\, e^m\, P_q \\
  &\qquad- 2[P^\sharp_c, P^\sharp_d]^q\, e^c \,e^d\, Q_q - 2 [Q^\sharp_c, Q^\sharp_d]^q \,e^c \,e^d \,P_q + \half N_\Jsh{\,}_{cd}{}^e \,e^c \,e^d \,e_e~.
\end{split} 
 \eeq
 The first and last term are $N_J$ and $N_{\Jsh}$.
%
%

\section{Some examples: deformations of $\o$ and $\O$ within $\IX$}
\vskip-10pt
We now illustrate the  extensions of some natural tensors in special geometry and calculate their deformations. For this subsection only,  $X$ is a Calabi--Yau manifold and we are at the standard embedding. It is  a good check that the formalism here reproduces the known deformation theory of a \cym.

\subsection{The hermitian form $\o$}
\vskip-10pt
The hermitian form $\Iomega$ on $\IX$ is  
\beq\notag
\Iomega \= \o_{mn} e^m e^n + \o^\sharp_{ab} \dd y^a \dd y^b~.
\eeq
with its form is determined by the metric \eqref{eq:ccFMetric} on $\IX$.

The variation of the hermitian form $\o$ due to a variation $\dd y^\a$ is
\beq\notag
 \fD_\a\o \= e_\a(\o)-(\Deth c_\a{}^m)\,\o_{m}~,
\eeq
and when decomposed into  holomorphic type this yields
\beq\label{varomega}
 \begin{split}
\fD_\a\o^{(2,0)}~&=~ 0 ~,\\[5pt]
\fD_\a\o^{(1,1)}~&=~  \fD_\a\o_{\m\nb}\,e^\m e^\nb  ~,\quad ~ \fD_\a\o_{\m\nb}=e_\a(\o_{\m\nb})-(\del_\m c_\a{}^\tau)\,\o_{\tau\nb}~,\\[5pt]
 \fD_\a\o^{(0,2)}~&=~   i\,\D_{\a}{}^\n\, \o_\n  ~,
 \end{split}
\eeq
Under a small diffeomorphism \eqref{eq:smallGauge} $\d \o \= \d y^a D_a \o$ transforms in the following way
$$
\d \o  ~\to~ \d \o - \cL_{\e} \o \= \d \o - \e^m (\dd \o)_m - \dd(\e^m\,\o_m) \= \d \o + \dd \ve  ~, \qquad \ve ~= - \e^m \o_m~.
$$
where $\e^m$ is a vector and $\ve$ a 1-form. Harmonic gauge is when $\dd^\dag \d \o = 0$. This requires 
$\Dethsharp_\a \o^{(0,2)} = 0$ and $\Dethsharp_\a \o^{(1,1)}$ to be harmonic.  

\subsection{The holomorphic form $\O$}
\vskip-10pt
We define the holomorphic three form on $X$ to have an extension which is purely vertical\footnote{Of course there may be a three form for $\IX$ with horizontal components. In this work, they do not play a role and so we do not consider them. This is an example of an object which does not have a natural extension to the universal manifold $\IX$. } 
\beq\label{eq:Omega}
 \O\=\frac{1}{3!}\,\O_{\m\n\r}\,e^\m e^\n e^\r \=\frac{1}{3!}\,f(x,y)\,\e_{\m\n\r} e^\m e^\n e^\r ~,
\eeq
where $\e_{\m\n\r}$ is the constant antisymmetric symbol and the  function $f$ depends holomorphically on the coordinates. As $\star \O = - \ii \O$, it follows $\O$ that is $\dd$-harmonic. Supersymmetry implies that it is covariantly constant with respect to the Bismut connection $\nabla^\Bi \O = 0$. Decomposing according to holomorphic type yields two relations
\begin{equation}\notag
\begin{split}
 \nabla_\m^\Bi\O &\=\Big(\del_\m\log{\norm{\O}^2}-H_{\m\n}{}^{\n}\Big)\,\O \= 0~, \\[3pt]
 \nabla_{\mb}^\Bi\O &\=-g^{\n\lb}(\del_{\mb}g_{\lb\n}-\del_{\lb}g_{\mb\n})\,\O\hskip1pt 
 \= H_{\mb\lb}{}^{\lb}\ \O\=0 ~,
 \end{split}
\end{equation}
which are solved by
\begin{equation}\notag
 H_{\m\n}{}^\n \= 0~, \qquad \del_\m\log{\norm{\O}^2}\=0~.
\end{equation}
The three--form $\O$ is a section of a line bundle over the moduli space $M$ with a $\IC^*$--gauge symmetry
\begin{equation}\label{eq:linebundlesymmetry}
 \O\to\l(y)\,\O ~, \qquad \l \in \IC^*~.
\end{equation}

Now consider variations of $\O$, which need to be covariant under \eqref{eq:linebundlesymmetry}. The  derivative  $\dethsharp_\a \O$ is decomposed into holomorphic type on $X$:
\begin{equation}\notag
\dethsharp_\a\O=(\Dethsharp_\a\O)^{(3,0)}+(\Dethsharp_\a\O)^{(2,1)}~,
\end{equation}
where the superscripts refer to holomorphic type with respect to $J$.
Using $\{\Deth,\Dethsharp\}{\=}0$, applying $\Dethsharp$ to $\Deth \O {\=} 0$ and decomposing according to  
\begin{equation}\begin{split}
\dethb(\Dethsharp_\a\O)^{(2,1)}\=0~~~\text{and}~~~ 
\dethb(\Dethsharp_\a\O)^{(3,0)}+\deth(\Dethsharp_\a\O)^{(2,1)} \=0 ~.
\end{split}\label{eq:dOmega}
\end{equation}
The first equation defines a $\dethb$-closed form $\ch_\a\=\D_\a{}^\m\O_\m$.
For the second equation the Hodge decomposition of $(\Dethsharp_\a\O)^{(3,0)}$ with respect to the $\deth$-operator gives the sum of a harmonic form and a $\deth$-exact term. As $h^{3,0} = 1$, the harmonic term is $\O$ multiplied by  a parameter dependant coefficient $\ccK{}_{2\,\a}$
\begin{equation}\label{HodgedelO30}
 (\Dethsharp_\a\O)^{(3,0)}~=-\ccK{}_{2\,\a}\,\O+\deth\z_\a~.
\end{equation}
Multiplying this  equation by $\Ob$ and integrating over $X$, we see the coefficient $\ccK_{2\,\a}$  can be written as a derivative
\begin{equation}\notag
\ccK_{2\,\a} = \del_\a \ccK_2 ~; \qquad  \ccK_2 ~=- \log\left(\int_X\,\ii\,\O\,\overline\O\right)~.
\end{equation}
Under small diffeomorphisms, there is a transformation law for $\d \O = \d y^\a \Dethsharp_\a \O$ 
\begin{equation}\notag
\begin{split}
 & \d \O^{(3,0)} \to \d \O^{(3,0)}  -\del(\e{}^\m\,\O_\m)~~~\text{and}~~~\d \O^{(2,1)} \to \d \O^{(2,1)} -\delb(\e{}^\m\,\O_\m)~.
\end{split}
\end{equation}
Comparing this equation with \eqref{HodgedelO30} we see that we can solve  $\e{}^\m\,\O_\m= \d y^\a \z_\a$ with explicit solution given by 
$$
\ve^\n \= \frac{1}{2\norm{\O}^2}\,  \Ob^{\n\r\s} (\d y^\a \zeta_{\r\s} + (\del \xi^{(1,0)})_{\r\s})~,
$$
where $\xi^{(1,0)}$ is an arbitrary one form. With this choice  $\d \O^{(3,0)}$ is harmonic. We see that $\xi^{(1,0)}$ is a residual gauge freedom that does not affect $\d \O$.

%
%

Returning to the second equation in \eqref{eq:dOmega}, we see that it implies $\chi_\a$ is $\deth$-closed.

 The  derivative of $\O$ that is covariant both with respect the symmetry \eqref{eq:linebundlesymmetry} and  diffeomorphisms~is
\beq\notag
\fD_\a\O\= \big(\Dethsharp_\a+\ccK{}_{2\,\a}\big)\,\O \= \chi_\a = \D_\a{}^\m \O_\m~.
\eeq

%
%
\vskip50pt
\raggedright
\baselineskip=10pt
\bibliographystyle{utphys}
\bibliography{BibliographyUniversalGeomPC}

\begin{thebibliography}{10}
\bibitem{Hitchin:Gerbe}
N.~Hitchin, ``What is a {G}erbe?.'' Notices of the {A}merican {M}athematical
  {S}ociety.
\end{thebibliography}
\end{document}